\documentclass[twocolumn, tighten]{aastex62}
\usepackage{comment}
\usepackage{natbib}
\usepackage{CJK}

\def\etal{{et~al.}}

\def\HII{{H\,{\textsc{ii}}}}

\def\n2o2{{N2O2}}
\def\dd16{{N2S2 - N2H$\alpha$}}

\submitjournal{ApJ}

\shorttitle{Metallicity, $q$, and pressure variations of \HII\ regions in TYPHOON}
\shortauthors{Grasha \etal}

\begin{document}
\begin{CJK*}{UTF8}{gbsn}
\title{Metallicity, ionization parameter, and pressure variations of \HII\ regions in the TYPHOON spiral galaxies: NGC~1566, NGC~2835, NGC~3521, NGC~5068, NGC~5236, and NGC~7793} 

\author[0000-0002-3247-5321]{K. Grasha}
\email{kathryn.grasha@anu.edu.au}
\altaffiliation{ARC DECRA Fellow}
\affiliation{Research School of Astronomy and Astrophysics, Australian National University, Canberra, ACT 2611, Australia}   
\affiliation{ARC Centre of Excellence for All Sky Astrophysics in 3 Dimensions (ASTRO 3D), Australia}   

\author[0000-0002-4382-1090]{Q.H. Chen (陈千惠)}
\affiliation{Research School of Astronomy and Astrophysics, Australian National University, Canberra, ACT 2611, Australia}   
\affiliation{ARC Centre of Excellence for All Sky Astrophysics in 3 Dimensions (ASTRO 3D), Australia}   
\affiliation{CAS Key Laboratory for Research in Galaxies and Cosmology, Department of Astronomy, University of Science and Technology of China, Hefei 230026, China}
\affiliation{School of Astronomy and Space Sciences, University of Science and Technology of China, Hefei, 230026, China}

\author[0000-0003-4569-2285]{A.J. Battisti}
\affiliation{Research School of Astronomy and Astrophysics, Australian National University, Canberra, ACT 2611, Australia}   
\affiliation{ARC Centre of Excellence for All Sky Astrophysics in 3 Dimensions (ASTRO 3D), Australia}   

\author[0000-0003-4804-7142]{A. Acharyya}
\affiliation{Department of Physics \& Astronomy, Johns Hopkins University, 3400 N. Charles Street, Baltimore, MD 21218, USA}
\affiliation{Research School of Astronomy and Astrophysics, Australian National University, Canberra, ACT 2611, Australia}   
\affiliation{ARC Centre of Excellence for All Sky Astrophysics in 3 Dimensions (ASTRO 3D), Australia}   

\author[0000-0003-3569-4092]{S. Ridolfo}
\affiliation{Research School of Astronomy and Astrophysics, Australian National University, Canberra, ACT 2611, Australia}   
\affiliation{ARC Centre of Excellence for All Sky Astrophysics in 3 Dimensions (ASTRO 3D), Australia}   

\author[0000-0002-4395-562X]{E. Poehler}
\affiliation{University of Canterbury, 20 Kirkwood Avenue, Upper Riccarton, Christchurch, 8041, New Zealand}

\author[0000-0003-1351-7351]{S. Mably}
\affiliation{University of Canterbury, 20 Kirkwood Avenue, Upper Riccarton, Christchurch, 8041, New Zealand}

\author[0000-0003-2396-4569]{A.A. Verma}
\affiliation{University of Canterbury, 20 Kirkwood Avenue, Upper Riccarton, Christchurch, 8041, New Zealand}

\author{K.L. Hayward}
\affiliation{University of Canterbury, 20 Kirkwood Avenue, Upper Riccarton, Christchurch, 8041, New Zealand}

\author{A. Kharbanda}
\affiliation{University of Canterbury, 20 Kirkwood Avenue, Upper Riccarton, Christchurch, 8041, New Zealand}

\author[0000-0001-5843-132X]{H. Poetrodjojo}
\affiliation{Sydney Institute for Astronomy (SIfA), School of Physics, The University of Sydney, Sydney, NSW 2006, Australia}
\affiliation{ARC Centre of Excellence for All Sky Astrophysics in 3 Dimensions (ASTRO 3D), Australia}   

\author[0000-0002-1143-5515]{M. Seibert}
\affiliation{The Observatories, Carnegie Institution for Science, 813 Santa Barbara Street, Pasadena, CA 91106, USA}

\author[0000-0002-5807-5078]{J.A. Rich}
\affiliation{The Observatories, Carnegie Institution for Science, 813 Santa Barbara Street, Pasadena, CA 91106, USA}

\author[0000-0002-1576-1676]{B.F. Madore}
\affiliation{The Observatories, Carnegie Institution for Science, 813 Santa Barbara Street, Pasadena, CA 91106, USA}
\affiliation{Department of Astronomy and Astrophysics, University of Chicago, Chicago, IL, USA}

\author[0000-0001-8152-3943]{L.J. Kewley}
\affiliation{Research School of Astronomy and Astrophysics, Australian National University, Canberra, ACT 2611, Australia}   
\affiliation{ARC Centre of Excellence for All Sky Astrophysics in 3 Dimensions (ASTRO 3D), Australia}

\begin{abstract}
We present a spatially-resolved \HII\ region study of the gas-phase metallicity, ionization parameter, and ISM pressure maps of 6 local star-forming and face-on spiral galaxies from the TYPHOON program. Self-consistent metallicity, ionization parameter, and pressure maps are calculated simultaneously through an iterative process to provide useful measures of the local chemical abundance and its relation to localized ISM properties. We constrain the presence of azimuthal variations in metallicity by measuring the residual metallicity offset $\Delta$(O/H) after subtracting the linear fits to the radial metallicity profiles. We however find weak evidence of azimuthal variations in most of the galaxies, with small (mean 0.03~dex) scatter. The galaxies instead reveal that \HII\ regions with enhanced and reduced abundances are found distributed throughout the disk. While the spiral pattern plays a role in organizing the ISM, it alone does not establish the relatively uniform azimuthal variations we observe. Differences in the metal abundances are more likely driven by the strong correlations with the local physical conditions. We find a strong and positive correlation between the ionization parameter and the local abundances as measured by the relative metallicity offset $\Delta$(O/H), indicating a tight relationship between local physical conditions and their localized enrichment of the ISM. Additionally, we demonstrate the impact of unresolved observations on the measured ISM properties by rebinning the datacubes to simulate low-resolution (1~kpc) observations, typical of large IFU surveys. We find that the ionization parameter and ISM pressure diagnostics are impacted by the loss of resolution such that their measured values are larger relative to the measured values on sub-\HII\ region scales. 
\end{abstract}

\section{Introduction}\label{sec:intro}
The content of heavy elements in a galaxy is one of the key properties for understanding its formation and evolutionary history, illuminated by gas-phase metallicity abundance variations in the interstellar medium \citep[ISM;][]{ma16, torrey19, maiolino19}. The gas-phase metallicity sets the balance between processes that enrich gas, such as star formation, and processes that dilute or remove metals, such as inflows of pristine gas from the intergalactic medium, galactic winds, and outflows. The gas-phase metallicity strongly influences the spectral emission from \HII\ regions in galaxies. \HII\ regions trace recent star formation because they are fuelled by the ionising photons of young, massive stars that are responsible for chemical evolution in spiral galaxies \citep{henry99}. The gas-phase metallicity is usually calculated as the oxygen abundance relative to hydrogen in units of 12 + log(O/H). Oxygen is used to define the overall gas-phase metallicity because oxygen is the dominant element by mass in the universe and is readily observable in the optical spectrum using temperature-sensitive collisionally excited lines \citep{kewley19_araa}.

The concept of analyzing metallicity variations in galaxies began with the seminal work of \citet{aller42}, who performed the first observation of radially-decreasing optical emission line ratios of [OIII] in M33. Since then, robust measurements have quantified the metallicity gradients that are present in nearly all massive disk galaxies \cite[e.g.,][]{searle71, shields74, vila-costas92, zaritsky94, sanchez12, sanchez14, sanchez-menguiano18, poetrodjojo21}. Such observations now allow for theoretical galaxy chemical evolution models to place quantitative constraints on key physical processes that drive galaxy evolution \citep[e.g.,][]{tinsley80, edmunds95, kobayashi07, prantzos00, torrey12, taylor16, bellardini21, sharda21}.

\HII\ regions reflect and trace the current metal content of the local ISM, assumed to be the same as the most recent generation of short-lived OB stars that are the power sources ionising the \HII\ regions. However, the metal content of the ISM is the consequence of the full chemical enrichment history at the location in which the stars from and therefore, the \HII\ region is created. The characterized negative radial metallicity gradients in spiral galaxies is believed to reflect the inside-out formation history of galaxies \citep[e.g.,][]{boissier99, chiappini01, fu09, ho15, belfiore17, belfiore19}. Environmental factors can additionally affect metallicity gradients, in particular, interacting galaxies, inflows of pristine gas diluting oxygen abundance in the galactic disk, or efficient radial mixing of gas \citep{bresolin09, rupke10, kewley10, torrey12, rich12, rosa14}. 

Although less studied, azimuthal variations of oxygen abundance in spiral galaxies is also an area of interest. Spiral arms are structures of enhanced star formation, which in turn may affect the chemical composition of these star-forming structures and produce chemical variations between arm and interarm regions. Studies of azimuthal variations in the chemical abundance distribution of spiral galaxies have shown conflicting results. Some studies do not observe significant variations in the gas metallicity between arm and interarm regions \citep[e.g., ][]{martin96, bresolin11, li13} whereas some studies find faint, but distinct azimuthal variations \citep[e.g., ][]{ho17, ho18, vogt17, kreckel19}. The conflicting results are quite often limited by the low number of analyzed galaxies, providing insignificant statistical samples. Variations in the azimuthal distribution of the gas-phase metallicity abundance implies that the dynamics of the streaming motion of gas and spiral density waves affects the chemical enrichment of the ISM via sub-kiloparsec-scale mixing. These observations provide the opportunity to compare with theoretical predictions for ISM mixing \citep{roy95, grand16, baba16, krumholz18}. \citet{ho19} found azimuthal variations in electron temperature in the galaxy NGC~1672, and attributed it to azimuthal variances in metallicity. This provides evidence that azimuthal variations are not driven by discrepancies in strong-line calibrations \citep[e.g.,][]{kewley08, bresolin16, peimbert17} but reflect real changes of the physical properties of the ISM in star-forming spiral galaxies. 

With the complete spatial coverage delivered by integral field spectroscopy (IFS), recent studies have ushered in a new era of characterising oxygen abundance variations radially and azimuthally. IFS coverage is preferential to traditional long-slit and multi-object spectroscopy studies, as the latter usually pre-select bright \HII\ regions from narrowband images and provide a biased view of metallicity variations. However, the limited field of view (FoV) of most current IFS's (16'' diameter for SAMI, 32'' diameter for MaNGA) limits the disk coverage for very local galaxies, where we can study the ISM in high-resolution. The key requirements to correctly characterizing the variations of the oxygen abundances are moderate spatial and spectral resolution across entire star-forming disks at high sensitivity in nearby galaxies \citep[see][for a recent review]{sanchez20}.

In this paper, we measure the spatial variations of oxygen abundance in the nearby spiral galaxies using IFS data obtained in the TYPHOON/PrISM (Progressive Integral Step Method) Program. The large FoV coverage of the TYPHOON datacubes (18' on one side, corresponding to deprojected galactocentric coverage of up to 25~kpc of the star-forming disk at the distances of the galaxies) provide us with an unparalleled sample to study variations in oxygen abundance in individual \HII\ regions across the entire star-forming disks. With the full coverage across the entire star-forming disks of these galaxies, the high spatial-resolution of our observations from TYPHOON of 25--150~pc are well matched to the typical spatial scales of \HII\ regions \citep[10--200~pc][]{azimlu11}. In addition to constraining the properties of the ISM conditions in local galaxies with a focus on the gas-phase metallicity, this work includes constraints on the ISM pressure to provide an additional alternative measure of the ISM conditions in these galaxies. 

The ISM pressure is an important physical condition that governs the emission from \HII\ regions, as well as derived quantities, such as the gas-phase metallicity, ionization parameter, and star formation rates. The nebula pressure is driven by both the mechanical energy produced by the central stellar population and the strength and shape of the radiation field. Most pressure diagnostics depend strongly on the gas-phase metallicity, which is problematic due to large discrepancies of up to 1~dex in 12+log(O/H) present between different metallicity calibrations \citep[see][for a review and discussion]{kewley08}. Because of this, theoretical ISM pressure calibrations can only be used with metallicity calibrations that have been constructed using consistent theoretical models. New self-consistent photoionization models are now available, which incorporate the detailed temperature and complex ionization structure of nebula to allow us to measure the ISM pressure consistently with metallicity in realistic \HII\ regions. 
In this work, we utilize the new theoretical ISM pressure models \citep{kewley19_apj} that have been constructed using consistent theoretical models of gas-phase metallicity calibrations to shed light on the physical properties of the resolved ISM gas conditions. 

This paper is organized as follows. Section~\ref{sec:obs} gives an overview of the TYPHOON survey and the galaxies in this paper. Section \ref{sec:analysis} outlines the procedure taken to measure the resolved ISM properties in individual \HII\ regions. The \HII\ region physical conditions are discussed in Section~\ref{sec:phys_cond} and the results are outlined in Section~\ref{sec:results}. In Section~\ref{sec:disucssion} we discuss the findings in the context of previous and future work. Finally, a brief summary is given in Section~\ref{sec:conclusion}.

\section{Observations and data reduction}\label{sec:obs}

\begin{figure*}
\includegraphics[scale=0.68]{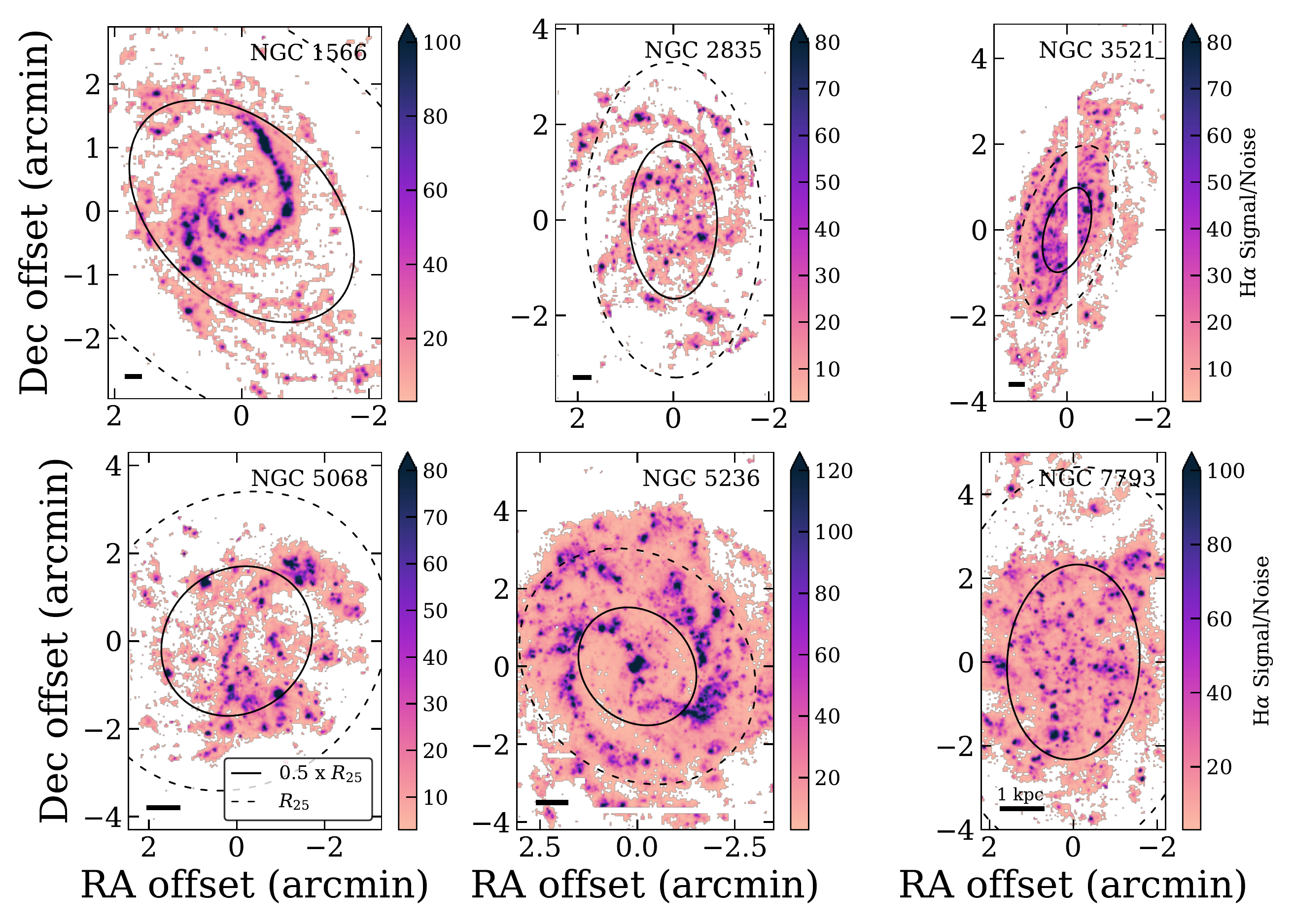}
\caption{H$\alpha$ signal to noise (S/N) distribution from fitting the emission line in the data cube for each of the TYPHOON galaxy in this study. Spaxels with S/N less than 3 are not shown. The solid and dashed ellipses correspond to $0.5 \times R_{25}$ and $R_{25}$ in the frame of the galactic disk, respectively. The black bar at the bottom left of each subplot represents 1~kpc at the distance of each galaxy.} 
\label{fig:galaxies}
\end{figure*}

\subsection{The TYPHOON Survey}
The six galaxies in this study --- NGC~1566, NGC~2835, NGC~3521, NGC~5068, NGC~5236, and NGC~7793 --- are drawn from the TYPHOON survey. The H$\alpha$ signal to noise image of each galaxy are shown in Figure~\ref{fig:galaxies}. The TYPHOON data are constructed using a Progressive Integral Step Method (PrISM), establishing a 3D data cube where each spatial pixel (spaxel) occupies a spectrum for that pixel. A long-slit spectrum is captured and then the slit is moved along spatially, generating multiple slices to form into a cube. Details about the TYPHOON Program and long-slit data reduction techniques will be presented in a forthcoming paper (M. Seibert et al., in preparation). Here, we provide a short summary, focusing on aspects immediately relevant to this study.

The TYPHOON galaxies (PI and Team Leader, B.F. Madore) were observed using the Wide-Field CCD imaging spectrograph on the 2.5m du~Pont telescope, located at the Las~Campanas Observatory in Chile. The WFCCD has a field of view of 25' ; the TYPHOON 3D datacubes are constructed using a custom long-slit (18' × 1.65; 0.5 square arcminute). The slit was positioned as either north-south or east-west, moved after each 600s to scan the data in the perpendicular direction of each galaxy. This was performed until the entire optical disk of each galaxy had been observed. The observations were taken over the course of observing runs from 2011 through 2018. Observations were performed only when the seeing was less than the slit width of 1$''$65. By choosing nearby galaxies ($z\leq0.005$), TYPHOON, and the galaxies in this study, is able to achieve seeing-limited resolutions of up to 5~pc. Standard spectro-photometric flux calibrations were taken each observing night.

The data are reduced using standard long-slit data reduction techniques, summarized in \citet{ho17} and to be described in detail in a forthcoming paper (M. Seibert et al., in preparation). We briefly summarize here the most important details. The wavelength calibration has a typical root-mean-square value 0.05~\AA\ for the entire data set. Flux calibration is accurate to $\sim$2\% at the spaxel scale over the range of 4500--7500~\AA. The reduced long-slit 2D spectra are tiled together to form the 3D data cube. The final reduced data cube covers a wavelength range of 3650--8150~\AA, with spectral and spatial samplings of 1.5~\AA\ and 1.65'', respectively. The instrumental dispersion is approximately $\sigma\sim3.5$~\AA, corresponding to a spectral resolution $R\sim850$ at 7000~\AA. We typically reach a 3$\sigma$ surface brightness sensitivity for H$\alpha$ of $4.2 \times 10^{-17}$~erg~s$^{-1}$~cm$^{-2}$~arcsec$^{-2}$. We note that as all main results in this paper (metallicity, ionization, and pressure diagnostics) are inferred from line ratios, the results depend only on the relative flux calibration and are insensitive to the absolute flux calibration.

\subsection{The TYPHOON Galaxies}\label{sec:galaxies}
The six galaxies selected for this study are all the spiral galaxies from the TYPHOON survey that have undergone the full data reduction pipeline at the time of writing, and additionally, were not previously published by \citet{ho17, ho18}. A summary of properties of the galaxies are listed in Table~\ref{tab:1} along with the spatial resolution of the observations and the number of \HII\ regions in each galaxy.

\subsubsection{NGC~1566}
The most distant galaxy in the sample at 17.9~Mpc, NGC~1566 is a face-on spiral galaxy with an intermediate-strength bar type, classified as a SAB(s)bc. NGC~1566 has a nuclear gas disk which presents as a star-forming ring \citep{smajic15} and an outer pseudo-ring that winds anti-parallel to the bar ends \citep{buta15}. NGC~1566 hosts a low-luminosity active galactic nuclei (AGN) that does not show evidence of an outflow nor AGN feedback \citep{combes14}. The moderate to loosely wound arms are consistent with being formed through bar-driven spiral density waves \citep{shabani18}. The largest \HII\ regions in NGC~1566 of $\sim$600~pc (Figure~\ref{fig:size}) are consistent with the maximum size of star-forming complexes of 508$\pm$179~pc as traced by the young stellar population \citep{grasha17b}.

\subsubsection{NGC~2835}
NGC~2835 is a highly inclined star-forming galaxy with a reported shallow radial metallicity gradient of $-0.096$~dex/kpc using the \citet{dopita16} metallicity calibration \citep{kreckel19}. The PHANGS-MUSE data from \citet{kreckel19} have a spatial resolution of 51~pc and thus are able to observe \HII\ regions that are 42\% smaller than the native resolution of the TYPHOON data. \citet{ryder95} use the $R_{23}$ diagnostic \citep{pagel79, kobulnicky04} to report a gradient of $-0.041$~dex/kpc. \citet{pilyugin14} used the data from \citet{ryder95} to provide homogeneous abundance determinations for \HII\ regions using T$_e$-based abundances, reporting a negative metallicity gradient of $-0.032$~dex/kpc.

\subsubsection{NGC~3521}
NGC~3521 is a highly inclined, flocculent spiral galaxy with a barred and inner ring morphology that hosts a stellar counter-rotating component induced by the bar \citep{zeilinger01}. There is an \HII\ region at the core of the nucleus that forms a low-ionization nuclear emission-line region, classified as an \HII\ LINER \citep{das03}.

\citet{pilyugin14} reports a negative metallicity gradient of $-0.039$~dex/kpc using T$_e$-based abundances based less than 12 individual \HII\ regions from \citet{zaritsky94, bresolin99}.

\subsubsection{NGC~5068}
NGC~5068 is a barred spiral galaxy with an established population of Wolf-Rayet stars --- helium-burning stars that are descendants of massive O stars with very strong stellar winds. Wolf-Rayet stars, with their spectroscopic signatures, are often serendipitously detected in observations of bright \HII\ regions \citep{rosa86}. From a spectroscopic survey of the WR population in NGC~5068, \citet{bibby12} derive a metallicity gradient of $-0.11$~dex/kpc using the \citet{pettini04} metallicity calibration. Using the $R_{23}$ metallicity calibration with observations from the Siding Spring Observatory, \citet{ryder95} report a gas-phase metallicity gradient of $-0.046$~dex/kpc. \citet{ryder95} additionally report a negative relationship between the oxygen abundance and surface brightness of the \HII\ regions. \citet{pilyugin14} used the data from \citet{ryder95} to provide homogeneous abundance determinations for \HII\ regions using T$_e$-based abundances, reporting a negative metallicity gradient of $-0.062$~dex/kpc.

\subsubsection{NGC~5236}
NGC~5236 (M83) is a face-on, weak-barred grand-design spiral that likely interacted with the nearby peculiar dwarf galaxy NGC~5253 within the last billion years, resulting in starburst activity in the central regions of both galaxies \citep{calzetti99}. 

A shallow radial metallicity gradient of $-$0.043~dex/kpc has been previously reported \citep{bresolin02} using the strong line diagnostics of \citet{kobulnicky99} with a flattening in the metallicity gradient occurring beyond the $R_{25}$ isophotal radius \citep{bresolin09, pilyugin12}. Using \emph{Spitzer Space Telescope} observations, \citet{dong08} found the presence of young star clusters in the outer regions of NGC~5236 at distances of up to $\sim$20~kpc from the galactic center, suggesting that the outer disk should contain a moderately chemically-evolved ISM. Using a compilation of data from the literature \citep{dufour80, bresolin99, bresolin02, bresolin05, esteban09, bresolin09}, \citet{pilyugin14} provide homogeneous abundance determinations for using a T$_e$-based abundances, reporting a flat negative metallicity gradient of $-0.0256$~dex/kpc.

At the resolution scales of TYPHOON, \citet{poetrodjojo19} resolved individual \HII\ regions without any diffused ionized gas (DIG) contamination and demonstrated that DIG contamination caused gradient smoothing at kiloparsec resolution scales because the light from high metallicity regions mixed with that from regions of lower metallicity.

\subsubsection{NGC~7793}
NGC~7793 is the closest galaxy in the sample, located at a distance of 3.62 +/- 0.15~Mpc \citep{jacobs09, anand21}. Resolved stars reveal a radial profile that exhibits a break at 5.1~kpc with the younger stellar populations exhibiting a steeper profile beyond the break \citep{radburn-smith12}. This is indicative of high levels of stellar radial migration and supports inside-out growth \citep{sacchi19}. 

From Wolf-Rayet stars, \citet{bibby10} estimated a metallicity gradient of $-0.078$~dex/kpc based on strong-line calibrations, consistent with the reported metallicity gradient inside the $R_{25}$ isophotal radius by \citet{stanghellini15}. Using a compilation of data from the literature \citep{webster83, edmunds84, mccall85, bibby10}, \citet{pilyugin14} provide homogeneous abundance determinations for using a T$_e$-based abundances, reporting a negative metallicity gradient of $-0.0662$~dex/kpc. The largest \HII\ regions in NGC~7793 of $\sim$200~pc (Figure~\ref{fig:size}) are consistent with the maximum size of star-forming complexes of 203$\pm$30~pc as traced by the young stellar population \citep{grasha17b, grasha18}.

\begin{deluxetable*}{lccccccccc}
\tabletypesize{\scriptsize}
\tablecaption{Properties of the TYPHOON galaxies\label{tab:1}}
\tablecolumns{10}
\tablewidth{0pt}
\tablehead{
\colhead{Galaxy} 		& 
\colhead{Distance} 	& 
\colhead{RC3 Radius R$_{25}$} & 
\colhead{log M$_\star$} & 
\colhead{log SFR} 	& 
\colhead{Inclination} & 
\colhead{P.A.} 			& 
\colhead{RC3 type} & 
\colhead{Resolution} &
\colhead{\HII\ Regions}
\\		
\colhead{} & 
\colhead{(Mpc)} & 
\colhead{(arcmin)} & 
\colhead{(M$_\odot$)} & 
\colhead{(M$_{\odot}$/yr)} & 
\colhead{(degree)} & 
\colhead{(degree)} & 
\colhead{} & 
\colhead{(pc/px)} &
\colhead{}
}
\colnumbers
\startdata 
NGC 1566 & 17.7  & 4.16	& 10.67 	& 0.65     	& 49 	& 44 	& SAB(s)bc	& 142 	& 166 \\ 
NGC 2835 & 12.2	& 3.30 	& 9.67  	& $-$0.10	& 56 	& $-$91& SB(rs)c   	& 98 	& 130 \\ 
NGC 3521 & 13.2 & 2.04 	& 10.83	& 0.42      & 60 	& 107  	& SAB(rs)bc & 104 	& 92 \\ 
NGC 5068 & 5.2	& 3.62 	& 9.36  	& $-$0.56	& 27 	& 137  	& SAB(rs)cd & 42 	& 133 \\ 
NGC 5236 & 4.89	& 3.28 	& 10.41 	& 0.62     	& 2  		& 45   	& SAB(s)c   	& 39 	& 148 \\ 
NGC 7793 & 3.62	& 4.66 	& 9.25 		& $-$0.60 & 47 	& 94   	& SA(s)d    	& 29 	& 243 
\enddata
\tablecomments{
Columns list the 
(1) galaxy name; 
(2) distance (Mpc) from \citet{anand21} and references therein; 
(3) $R_{25}$ Mean radius of 25 mag arcsec$^{-2}$ isophote (B-band) from RC3 \citep{devaucouleurs91} in arcmin; 
(4) log of the stellar masses M$_\star$ (M$_\odot$) from \citet{leroy19}; 
(5) log of the star formation rate (SFR) M$_{\odot}$/yr from \citet{leroy19}; 
(6) galaxy inclination (degrees) from Vizier; 
(7) position angle measured from North to East measured on the V-band image;
(8) de Vaucouleurs morphology class \citep{devaucouleurs91};
(9) deprojected physical resolution (parsec/pixel); and 
(10) total number of \HII\ regions (Section~\ref{sec:HIIphot}).
}
\end{deluxetable*}

\section{Data analysis}\label{sec:analysis}
To constrain the distribution of the gas-phase oxygen abundance in the six TYPHOON galaxies, we first construct an H$\alpha$ emission line map from the reduced TYPHOON data cubes. We then identify the \HII\ regions in the H$\alpha$ maps using the code \texttt{HIIphot}. The composite emission line fluxes of each \HII\ region are measured from the 2D \texttt{HIIphot} \HII\ region masks to constrain the properties of the individual \HII\ regions.

\subsection{Identifying \HII\ Region Candidates with \texttt{HIIphot}}\label{sec:HIIphot}
The emission line fitting from \texttt{LZIFU} (Section~\ref{sec:linefluxes}) is first performed on every pixel in the datacube to construct the H$\alpha$ map. From the H$\alpha$ map, the \HII\ regions are identified using the IDL routine \texttt{HIIphot} \citep{thilker02}. \texttt{HIIphot} morphologically isolates \HII\ regions from the DIG on the narrowband H$\alpha$ image map and provides a 2D mask that identifies separate regions within the map. \texttt{HIIphot} has been successfully applied to narrowband images and IFS surveys of nearby galaxies to characterize physical properties of \HII\ regions \citep[e.g.,][]{thilker02, helmboldt05, ho17, kreckel19}. Table~\ref{tab:1} lists the number of \HII\ regions in the H$\alpha$ map for each galaxy based on the H$\alpha$ surface brightness distribution that pass our selection criteria (Section~\ref{sec:criteria}).

\texttt{HIIphot} works by finding areas of maximum H$\alpha$ flux, a tracer of ionized gas characteristic of star-forming regions. These maxima act as `seeds' as the initial guess for the \HII\ region. The seeds are then iteratively grown until stopped by a pre-set termination condition. This termination condition is one of several that are user-specified. The boundary of an \HII\ region is determined by the gradient of the H$\alpha$ surface brightness where the growth for a particular region stops when the observed surface brightness profile flattens sufficiently and/or no more qualified pixels can be reached owing to being surrounded by other regions. 

The settings of \texttt{HIIphot} are adjusted to produce realistic \HII\ region sizes and shapes across the six galaxies in this sample. We then spatially bin the datacubes based on the \HII\ region mask as identified by \texttt{HIIphot}, to produce composite \HII\ region spectra. We then re-measure the emission line fluxes using \texttt{LZIFU} (Section~\ref{sec:linefluxes}). We use the measured emission lines from the composite \HII\ regions to report the flux ratios and physical properties throughout this paper. We find broad agreement between the derived line ratios and metallicities of the \HII\ regions calculated from the \texttt{HIIphot} \HII\ region catalogs versus extracting spectra from a circular 1$''$ diameter aperture centered at the position of each identified \HII\ region. This confirms that the exact boundaries from \texttt{HIIphot} and the tuning parameters have a relatively minor impact on the resulting \HII\ region masks and that \texttt{HIIphot} performs well at separating the \HII\ region flux from contamination via DIG emission. This additionally supports the validity of measuring metallicity gradients with longslit and mask observations.

\begin{figure*}
\includegraphics[scale=0.56]{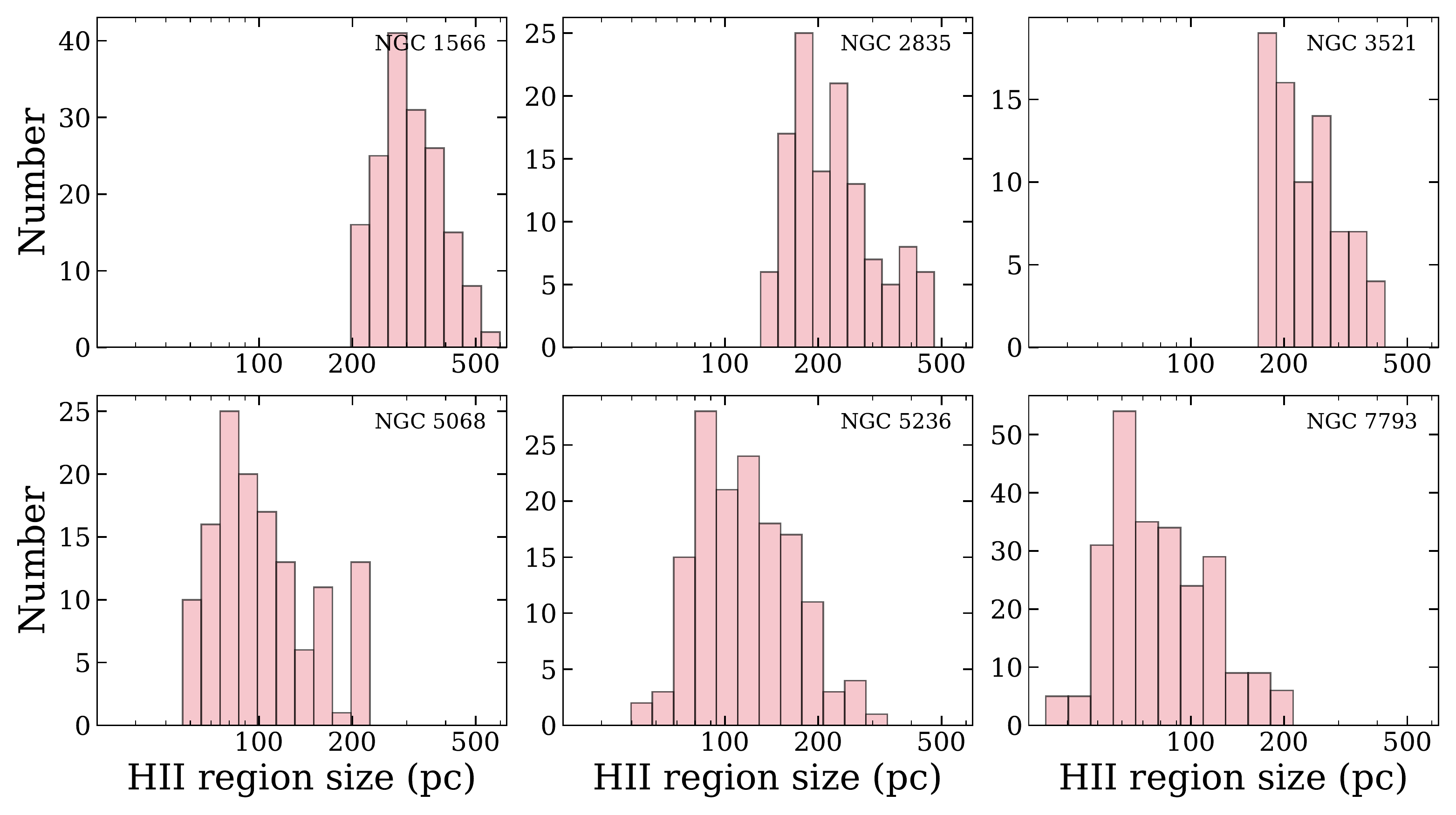}
\caption{Distribution of the \HII\ region sizes (radius) from \texttt{HIIphot}, for each galaxy.}
\label{fig:size}
\end{figure*}

Figure~\ref{fig:size} shows the sizes of the \HII\ regions which are computed as the circular radius that contains the same area as covered by the \HII\ region masks. The size distribution exhibits the expected power-law behavior observed in \HII\ regions \citep[e.g., ][]{kreckel19} modulo small number statistics within each galaxy. In general, Figure~\ref{fig:size} shows that there is a strong dependency on the size of the \HII\ region and distance of the galaxy, where the furthest galaxies in our sample result in \HII\ regions that are consistent with sizes of unresolved large star-forming complexes that are composed of multiple, unresolved individual \HII\ regions \citep{grasha17a, grasha17b}. 

It is important to note that identifying \HII\ regions is non-trivial, particularly in the center of galaxies due to source crowding. The primary purpose of creating \HII\ region masks and synthesizing the composite \HII\ region spectra is to increase the S/N of the weak lines as well as to avoid line fluxes that are contaminated by non-thermal emission (e.g. diffuse ionized gas, shocked gas, AGNs). It is possible that a single \HII\ region that we have defined in our galaxies may in-fact be a composite of multiple, poorly resolved individual \HII\ regions. We note that in this case, the oxygen abundance and ionization parameter derived using the integrated line fluxes are approximately the luminosity-weighted mean oxygen abundance and ionization parameter.

\subsection{Emission Line Fluxes}\label{sec:linefluxes}
We measure the emission line fluxes of each composite \HII\ region using \texttt{LZIFU} \citep[LaZy-IFU;][]{ho16}. \texttt{LZIFU} is an emission line fitting tool created for processing integral field spectroscopy (IFS) data. \texttt{LZIFU} uses \texttt{pPXF} \citep{cappellari04, cappellari17} to model and subtract the stellar continuum for each spaxel (or in our case, the composite \HII\ region flux) from the emission line component. The emission lines are fitted as Gaussian components using the Levenberg--Marquardt least-squares method. Following the procedure of \citet{ho17}, we fit the stellar continuum using the MIUSCAT single-burst stellar solar-scaled theoretical simple stellar population models \citep{vazdekis12} with 13 ages (logarithmic scale between 0.063 and 15.85~Gyr) and 3 metallicities ([Fe/H] = -0.71, 0, 0.22). 

The emission lines we fit as part of this study are [OII]$\lambda\lambda 3727, 3729$, H$\beta$, [OIII]$\lambda\lambda 4959, 5007$, H$\alpha$, [NII]$\lambda\lambda 6548, 6583$, and [SII]$\lambda\lambda 6717, 6730$. For this study, we only fit a single Gaussian component to each emission line. The 2D map of emission line fluxes and errors for each region are extracted and used for further analysis. The velocities and velocity dispersions of all the lines are constrained to a single velocity and tied to each other and the flux ratios of [OIII]$\lambda\lambda4959, 5007$ and [NII]$\lambda\lambda6548, 6583$ are fixed to the ratio given by quantum mechanics as their ratios are independent of physical conditions \citep{gurzadyan97}. \texttt{LZIFU} returns maps of the flux and flux errors for each emission line, as well as maps of the ionized gas velocity and velocity dispersion and their associated errors \citep[see][for a detailed explanation of the routine]{ho16}. 

We propagate line flux errors produced by LZIFU through to the metallicity calculations for the composite \HII\ regions following the method of \citet{bianco16}. This method uses the Monte Carlo sampling to better characterize the statistical oxygen abundance confidence region. We use the line flux measurements and their uncertainties as inputs to simulate 2000 maps for all emission lines used in the calculation to provide estimates for the 68\% confidence regions. As this method does not include systematic uncertainties, we also incorporate the systematic uncertainties in the calibration and reddening correction. This measure of the errors of the maps are propagated to the gradient errors.

\subsection{Flux Extinction and Reddening Correction}\label{sec:reddening}
As the attenuation of emission lines is wavelength-dependent, it is imperative to correct emission line fluxes for the effect of dust extinction, especially for diagnostics that utilize emission lines separated by large a wavelength range such as the \n2o2\ metallicity diagnostic. We calculate the Balmer ratio (H$\alpha$/H$\beta$)$_{\rm observed}$ to solve for the color excess $E(B-V)$ using the relationship
\begin{equation}
E(B-V) = \frac{\log \left(\frac{(\mathrm{H}\alpha/\mathrm{H}\beta)_{\rm observed}}{(\mathrm{H}\alpha/\mathrm{H}\beta)_{\rm intrinsic}}\right)}{0.4 \times \left(k(\mathrm{H}\beta) - k(\mathrm{H}\alpha)\right)},
\end{equation}
where $(\mathrm{H}\alpha/\mathrm{H}\beta)_{\rm intrinsic}$ is the intrinsic ratio of 2.86$^{+0.18}_{-0.11}$ for case B recombination \citep{osterbrock89} for temperatures of T$_e$ = 10,000$^{+10,000}_{-5000}$~K at a density of 100~cm$^{-3}$ and $k(\mathrm{H}\beta)$ and $k(\mathrm{H}\alpha)$ are the values of the extinction curve at the wavelengths of H$\beta$ and H$\alpha$, respectively. We use the Milky Way extinction curve $k(\lambda)$ of \citet{fitzpatrick99} and assume a typical R(V) value of 3.1 to determine the extinction at the wavelength of each emission line ($k(\lambda)=A(\lambda)/E(B-V)$). We correct the flux of all emission lines for extinction to recover the intrinsic, un-reddened fluxes.

\subsection{Star-Forming \HII\ Region Selection Criteria}\label{sec:criteria}
Ratios of emission lines can shed light on the dominant ionizing mechanism. \citet{baldwin81} (hereafter BPT) first pioneered the method of comparing two sets of emission line ratios to distinguish between star-forming galaxies and AGN, further developed by \citet{veilleux87}. Following the birth of this method, it is now commonplace to determine \HII\ region types by analyzing their location on a BPT diagram. Additions from \citet{kewley01} and \citet{kauffmann03} have made this method more straightforward with modeled lines to delineate the upper and lower limits of each region in a BPT diagram. The line defined in \citet{kewley01} was created using photoionization and stellar population synthesis models and gave a lower bound on AGN types in a sample. Any objects found above this line cannot fit the star-forming model. We use the original delineation models from \citet{kewley01} and \citet{kauffmann03} to identify and separate spaxels with non-stellar contributions; a future paper will focus on deriving new BPT relationships from the updated models.

After \HII\ regions were found using \texttt{HIIphot} (Section~\ref{sec:HIIphot}), we additionally exclude regions with line ratios that are inconsistent with photoionization and might be influenced by AGN and/or shocks based on their [NII]/H$\alpha$ and [OIII]/H$\beta$ line ratios \citep{kewley01, kauffmann03, kewley06}. To determine the dominating excitation source for each \HII\ region, we use the classification scheme of \citet{kauffmann03} to distinguish when non-star-forming emission is present using the following strong emission line ratio diagnostic curve of the BPT diagram:
\begin{equation}
\log \left( \frac{[O III]}{H\beta}\right)> \frac{0.61}{\log([NII]/H\alpha)-0.05} + 1.3.
\end{equation}
The \citet{kauffmann03} line represents the divide in which \HII\ regions are arising due to thermal or non-thermal sources; \HII\ regions that lay above this demarcation are rejected from all analysis in this study. 

Lastly, we place further selection cuts and exclude \HII\ regions with signal to noise ratios below 3 in the strong emission lines of H$\alpha$, H$\beta$, [OIII], [NII], and [SII]. There are a total of 21 regions that do not make the required S/N cut or show line ratios inconsistent with photoionization (69 regions). The lower throughput in the blue part of the spectrograph causes the S/N of the [OII] emission line to be relatively low compared to the other optical emission lines and as a result, we do not place a S/N cut on the [OII] line. However, because we require the [OII] emission lines to calculate metallicity and ionization parameter, we exclude any \HII\ regions which do not have reported [OII] emission as measured by \texttt{LZIFU}. \HII\ regions that satisfy all these criteria are classified as star-forming regions and make up the final selection of \HII\ regions in these galaxies (Table~\ref{tab:1}).

\section{\HII\ region physical conditions}\label{sec:phys_cond}

\subsection{Gas-phase Oxygen Abundance \n2o2}\label{sec:n2o2}
In this work, we derive the oxygen abundances for the \HII\ regions from the optical data cubes using the popular optical metallicity diagnostic \n2o2\ that uses the brightest nitrogen and oxygen emission lines, [NII]$\lambda$6583/[OII]$\lambda\lambda$3726,3729 \citep{kewley02}. 
The \n2o2\ metallicity diagnostic is a reliable metallicity diagnostic in the optical spectrum \citep{kewley19_araa}, advantageous because the [NII] and [OII] lines are unaffected by underlying stellar absorption. In addition, the similar ionizing potentials of the nitrogen and oxygen species result in a diagnostic with little dependence on the ionization parameter and only marginal dependence on the ISM pressure. Lastly, the \n2o2 diagnostic is also minimally affected by DIG contamination \citep{zhang17}. DIG contamination is not an issue we need to worry about because HIIphot performs well in separating out \HII\ region from DIG emission. The primary downside of the \n2o2\ diagnostic is the strong dependence on extinction due to the large wavelength differences between the [NII] and [OII] emission lines. In addition, the \n2o2\ diagnostic in unreliable at low metallicities where the nitrogen to $\alpha$-elements abundance ratio (N/$\alpha$) with O/H flattens, however, this does not impact the galaxies in our sample. In this work, we implement the \n2o2 metallicity diagnostic as parametrized in \citet{kewley19_araa} for an ISM pressure of log($P/k$) = 5.0. 

The gas-phase oxygen abundance can be derived from numerous diagnostics and calibrations that are available in the literature. Despite the different diagnostics yielding discrepant oxygen abundance measurements \citep[see][]{kewley08}, these differences primarily impact the absolute metallicity values whereas the relative values of metallicity are often statistically robust between different calibrations \citep[e.g.,][]{ho17, poetrodjojo21}. As a result, in this work we do not attempt to compare the results derived from multiple calibrations and we only present the oxygen abundances derived from the \n2o2\ metallicity diagnostic \citep{kewley19_araa}. Thus, the main results of this work are independent of the adopted metallicity calibration as we are primarily concerned with the relative values of metallicity (see Section~\ref{sec:delOH}) and not absolute metallicity values.

\begin{figure*}
\includegraphics[scale=0.7]{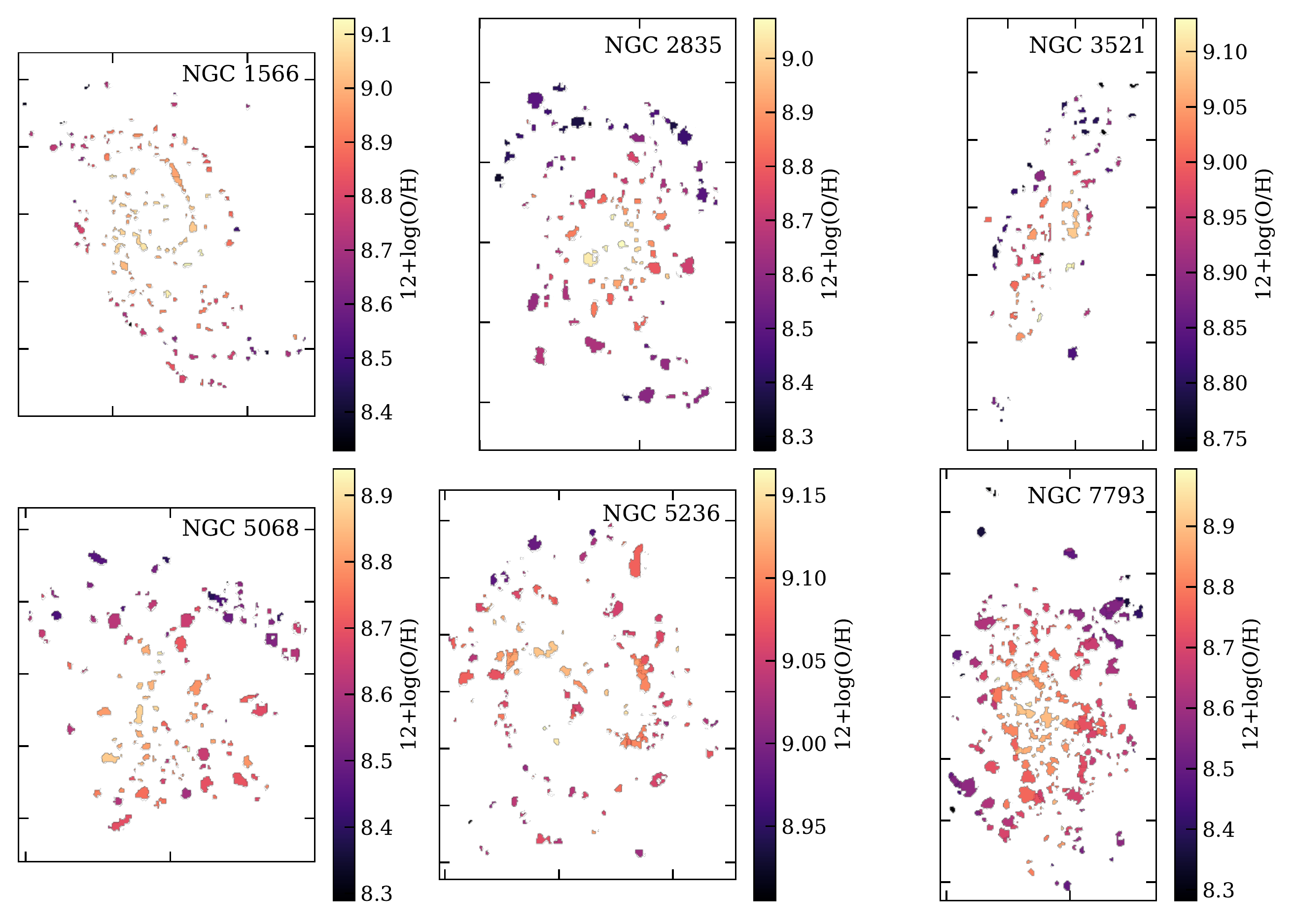} 
\caption{Gas-phase oxygen abundance maps using the \n2o2\ diagnostic (Section~\ref{sec:n2o2}). Each region on the maps corresponds to one \HII\ region identified from the H$\alpha$ surface brightness map (Section~\ref{sec:HIIphot}; Figure~\ref{fig:galaxies}) using \texttt{HIIphot} \citep{thilker00}.}
\label{fig:hiiregions_metallicity_n2o2}
\end{figure*}

Figure~\ref{fig:hiiregions_metallicity_n2o2} maps the \n2o2\ gas-phase metallicity of each \HII\ region in each galaxy. These figures demonstrate both the different mean metallicity between galaxies \citep[linked to the mass--metallicity relation][]{lequeux79, tremonti04} and the well-known trend of metallicity decreasing with galactic radius --- reflecting the inside-out formation history of galactic disks \citep{boissier99}. 

The \n2o2\ gas-phase metallicity maps (Figure~\ref{fig:hiiregions_metallicity_n2o2}) prominently displays the typical negative metallicity gradient, where the metallicity of a galaxy decreases radially from the center commonly seen in local disk galaxies \citep[e.g., ][]{vila-costas92}. A first-order linear fit to the radial gas-phase metallicity gradient is performed in Section~\ref{sec:metgradients}. Notably, there is not obvious azimuthal variation to the metallicity distributions. This is further investigated in Section~\ref{sec:azimuthal}.

\subsection{Ionization Parameter}\label{sec:ionization}
The ionizing radiation field sets the structure of an \HII\ region, powered by the central star clusters. In an idealized case of a spherically symmetric \HII\ region in equilibrium, this is parametrized as the dimensionless ionization parameter $U$\footnote{the dimensional ionization parameter $q$ (cm s$^{-1}$) is related to the dimensionless ionization parameter $U$ by dividing by the speed of light, $U = q/c$.} as:
\begin{equation}
U = \frac{Q(H^0)}{4\pi R^2n_Hc},
\end{equation}
where $Q(H^0)$ is the number of hydrogen ionising photons (energy $>$13.6~eV) emitted per second in the local area, $n_H$ is the density of the hydrogen atoms, $R$ is the radius of the \HII\ region, and $c$ is the speed of light. Higher values of ionization parameter indicate a stronger ionizing source and/or a lower ionized gas density. We note that interpretation of the ionization parameter is limited by both degeneracies between the the shape of the ionizing continuum, mass of stellar populations, and the asymmetrical nature of \HII\ regions made obvious by nearby, resolved studies of \HII\ regions \citep[e.g.,][]{pellegrini12}. The global ionization parameter in galaxies is usually anti-correlated with the gas-phase metallicity such that low-metallicity galaxies show larger ionization parameters \citep{dopita86}. The ionising radiation field from a star is modified by the metallicity of its stellar atmosphere through line blanketing and rotation effects \citep{dopita06}. The stellar atmosphere metallicity is determined by the metallicity of the gas that the star formed from. This metallicity-ionization parameter relation, however, may start to break down for spatially resolved data; in samples of resolved, isolated \HII\ regions, the ionization parameter is not found to correlate with metallicity \citep{garnett97, dors11}. One additional caveat do not include prescriptions for the theoretical models for the ionization parameter grids is that the presence of very massive rotating stars and their underlying chemical abundance \citep{grasha21}. Lastly, binary systems can heavily dominate the emission of ionizing photons. \citet{gotberg19} demonstrated that stripped stars from binaries have harder ionizing spectra than massive single stars, increasing the ionization parameter and the production efficiency of hydrogen ionizing photons. 

The optical spectrum contains two strong-line ionization parameter diagnostics to constrain the ionization parameter from photoionization modeling: [OIII]$\lambda$5007/([OII]$\lambda$3726+[OII]$\lambda$3729) and ([SIII]$\lambda$9069 + [SIII]$\lambda$9532)/([SII]$\lambda$6716 + [SII]$\lambda$6731) \citep{kewley02}. We calculate the ionization parameter using the O32 diagnostic of \citet{kewley19_araa}, calculated in an iterative process using the \n2o2\ metallicity diagnostic as TYPHOON does not observe the red [SIII]$\lambda\lambda$9069,9532 lines. 

\begin{figure*}
\includegraphics[scale=0.7]{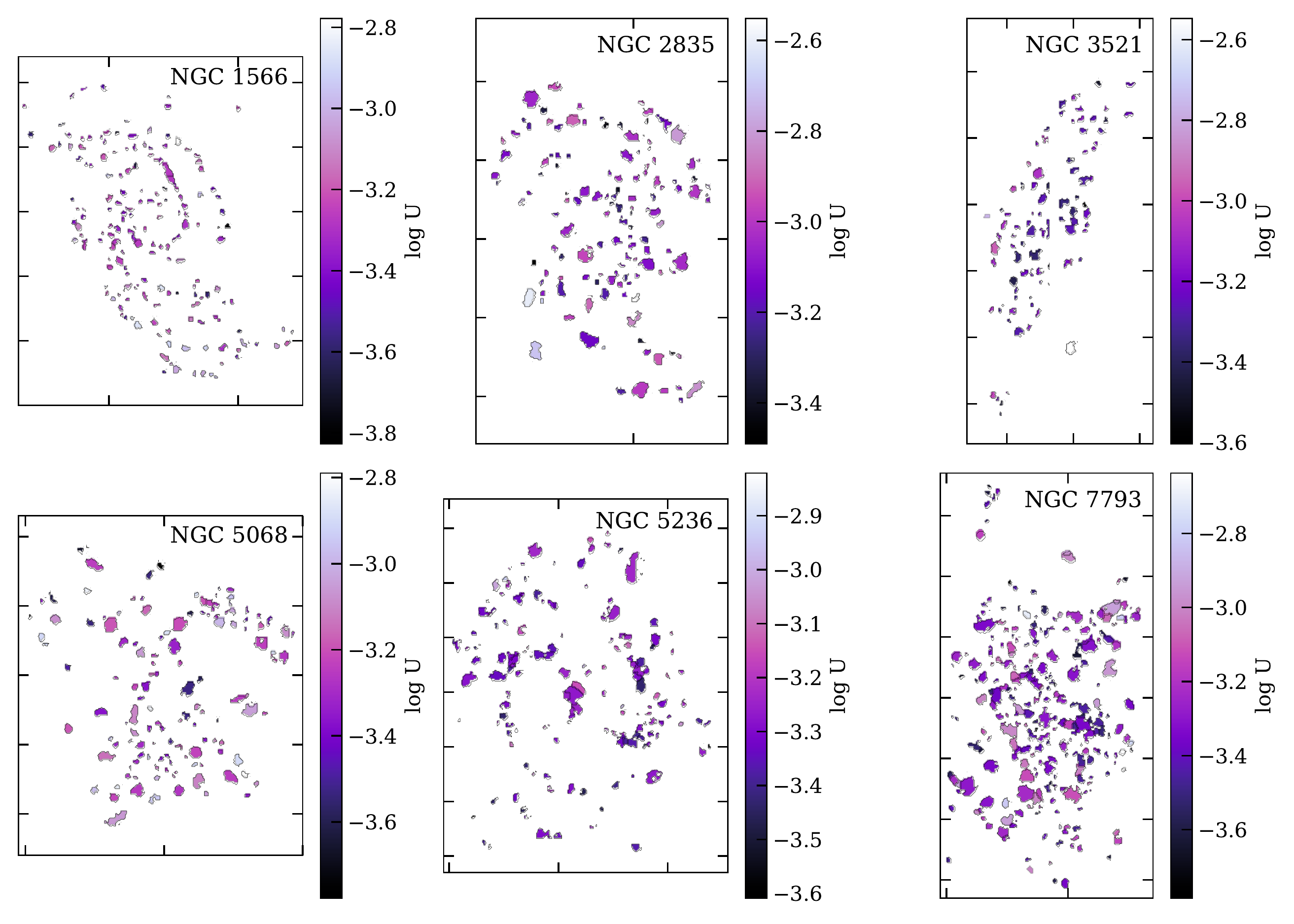}
\caption{Ionization parameter $U$ maps using the O32 diagnostic (Section~\ref{sec:ionization}). }
\label{fig:hiiregions_ionization}
\end{figure*}

Figure~\ref{fig:hiiregions_ionization} maps the O32 ionization parameter of each \HII\ region in each galaxy in this study. These figures demonstrate the remarkably uniform ionization parameter values across all the galaxies, consistent with the reported flat ionization parameter gradients in prior studies \citep{dopita00, kreckel19}. As highlighted by \citet{kewley19_apj}, different ionization lines of different species probe different zones within a nebula. Emission lines produced by different ions and different energy levels are sensitive to different density and temperature regimes. This in turn impacts the measured pressure as well (Section~\ref{sec:pressure}) as the pressure diagnostics of \citet{kewley19_apj} primarily depend on the gas-phase metallicity through its effect on the electron temperature of the gas. Thus, we encourage caution of direct comparison to studies which use observations of different ionization lines of different species to derive the ionization structure and pressure and to keep in mind the caveats this can cause (see further discussion in Section~\ref{sec:discussion_1}).

\subsection{ISM Pressure}\label{sec:pressure}
The nebular emission lines and derived quantities, such as the gas-phase metallicity, ionization parameter, and star formation rate, depend critically on assumptions about the pressure of a nebula. In this work, we use the new self-consistent theoretical optical diagnostics to measure the ISM pressure with the calibrations from \citet{kewley19_apj}. We measure the ISM pressure using the [SII]$\lambda\lambda 6717, 6730$ doublet line ratios. 

Because metals act as coolants in nebulae, the \textit{electron temperature}, too, is inherently linked to the gas-phase metallicity. The ISM pressure is governed by both the density and the temperature structure of the nebula. Therefore, ISM pressure diagnostics have a strong dependence on the gas-phase metallicity. In this work, since the density varies little for these galaxies and is consistent with the low-density limit, the ISM pressure is a more meaningful quantity to discuss than the electron temperature or ISM density \citep[see][]{kewley19_apj}.

The ISM pressure is measured by matching the [SII]6717/[SII]6730 emission line ratios to the closest values in the \citet{kewley19_apj} theoretical pressure model table, then interpolated linearly to find the corresponding pressure at the best-matched ionization parameter $U$ (Section~\ref{sec:ionization}) and gas-phase metallicity value from the \n2o2\ calibration (Section~\ref{sec:n2o2}). All pressure values are reported as log($P/k$) with units of cm$^{-3}$~K. 17\% of \HII\ regions across the sample show [SII]6717/[SII]6730 emission line ratios that are outside of the range of the model grid (corresponding to values of $4<\log(P/k)<9$). These \HII regions are excluded from all analyses related to the ISM pressure.

\begin{figure*}
\includegraphics[scale=0.7]{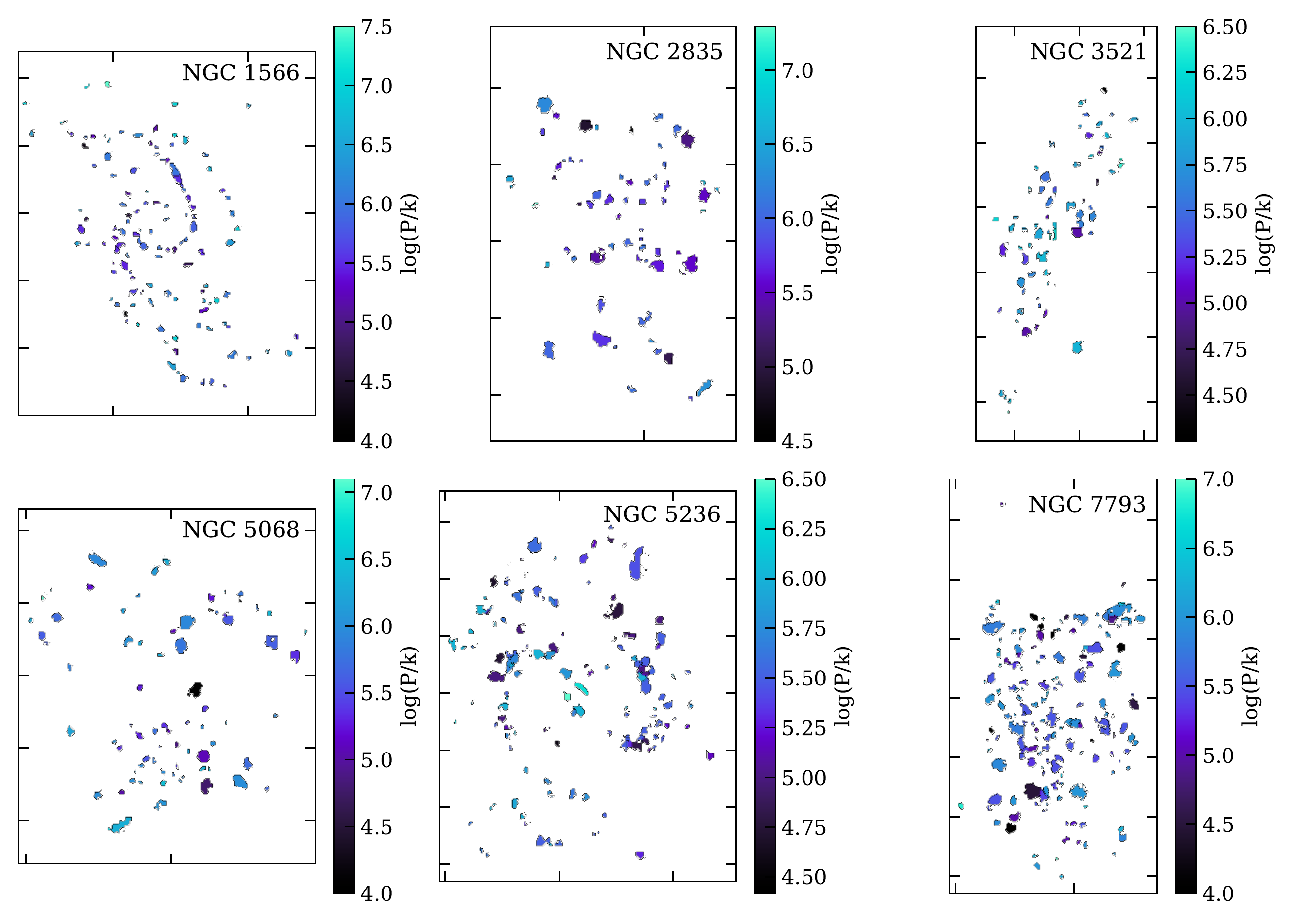}
\caption{Pressure maps using the [SII]6717/[SII]6730 emission line ratios (Section~\ref{sec:pressure}). }
\label{fig:hiiregions_pressure}
\end{figure*}

Figure~\ref{fig:hiiregions_pressure} maps the [SII]6717/[SII]6730 (S$_2$) pressure diagnostic of each \HII\ region in each galaxy in this study. Similar to ionization parameter (Figure~\ref{fig:hiiregions_ionization}), these maps demonstrate the rather remarkably uniform \HII\ region pressures across all the galaxies.

\movetabledown=4cm
\begin{rotatetable*}
\begin{deluxetable*}{lccccccccccc}
\tabletypesize{\scriptsize}
\tablecaption{Radial gradients of the \HII\ regions \label{tab:gradients_hii}}
\tablecolumns{12}
\tablewidth{700pt}
\tablehead{
\colhead{}  & 
\multicolumn{3}{c}{Gas-phase metallicity} & 
\colhead{}  & 
\multicolumn{3}{c}{Ionization parameter}  & 
\colhead{} & 
\multicolumn{3}{c}{Pressure}
\\		
\colhead{} & 
\multicolumn{3}{c}{\n2o2 \citep{kewley19_araa} -- Figure~\ref{fig:metgrad_n202}} & 
\colhead{} & 
\multicolumn{3}{c}{O32 \citep{kewley19_araa} -- Figure~\ref{fig:qgrad}}  & 
\colhead{} & 
\multicolumn{3}{c}{S2 \citep{kewley19_apj} -- Figure~\ref{fig:presgrad}}
\\
\cline{2-4} \cline{6-8} \cline{10-12}\noalign{\smallskip} 
\colhead{Galaxy} & 
\colhead{Central value} & 
\colhead{Gradient} & 
\colhead{Gradient} &
\colhead{} & 
\colhead{Central value}  & 
\colhead{Gradient} & 
\colhead{Gradient} &
\colhead{} & 
\colhead{Central value}  & 
\colhead{Gradient} & 
\colhead{Gradient} 
\\
\colhead{} & 
\colhead{(log(O/H)+12)} & 
\colhead{(dex/kpc)} & 
\colhead{(dex/R$_{25}$)} &
\colhead{} & 
\colhead{(log $U$)} & 
\colhead{(dex/kpc)} & 
\colhead{(dex/R$_{25}$)} &
\colhead{} & 
\colhead{(log P/k)} & 
\colhead{(dex/kpc)} & 
\colhead{(dex/R$_{25}$)} 
\\
}
\startdata 
NGC 1566	 & $9.14\pm0.02$		& $-0.0252\pm0.0018$ & $-0.54\pm0.04$	&& $-3.25\pm0.04$	& $0.004\pm0.003$	& $0.09\pm0.08$		&& $5.55\pm0.12$ & $0.041\pm0.011$ & $0.9\pm0.2$\\
NGC 2835	& $8.99\pm0.03$	    & $-0.043\pm0.003$	 	& $-0.50\pm0.04$		&& $-3.11\pm0.04$	& $0.003\pm0.005$	& $0.04\pm0.06$		&& $5.66\pm0.14$  & $0.03\pm0.02$  & $0.4\pm0.2$ \\
NGC 3521	& $9.076\pm0.018$  & $-0.020\pm0.003$	& $-0.16\pm0.02$		&& $-3.21\pm0.04$	& $0.007\pm0.006$	& $0.005\pm0.007$	&& $5.67\pm0.12$ & $0.002\pm0.016$ & $0.016\pm0.13$ \\
NGC 5068	& $8.88\pm0.02$	    & $-0.073\pm0.008$	 	& $-0.40\pm0.04$		&& $-3.23\pm0.05$	& $0.002\pm0.015$	& $0.011\pm0.08$	&& $5.29\pm0.19$ & $0.17\pm0.08$   & $0.9\pm0.4$ \\
NGC 5236	& $9.131\pm0.008$ 	& $-0.016\pm0.002$	 & $-0.075\pm0.008$ && $-3.19\pm0.03$	& $-0.005\pm0.008$& $-0.02\pm0.04$ 		&& $5.54\pm0.10$ & $-0.008\pm0.025$ & $0.04\pm0.13$ \\
NGC 7793	& $8.945\pm0.013$	& $-0.083\pm0.005$	 & $-0.41\pm0.02$	&& $-3.33\pm0.03$	& $0.015\pm0.010$	& $0.07\pm0.05$			&& $5.65\pm0.11$ & $0.02\pm0.04$   & $0.09\pm0.18$ \\
\enddata
\end{deluxetable*} 
\end{rotatetable*}

\movetabledown=4cm
\begin{rotatetable*}
\begin{deluxetable*}{lccccccccccc}
\tabletypesize{\scriptsize}
\tablecaption{Radial gradients of the 1~kpc spaxels \label{tab:gradient_1kpc}}
\tablewidth{0pt}
\tablehead{
\colhead{}  & 
\multicolumn{3}{c}{Gas-phase metallicity} & 
\colhead{}  & 
\multicolumn{3}{c}{Ionization parameter}  & 
\colhead{} & 
\multicolumn{3}{c}{Pressure}
\\		
\colhead{} & 
\multicolumn{3}{c}{\n2o2 \citep{kewley19_araa} -- Figure~\ref{fig:metgrad_n202}} & 
\colhead{} & 
\multicolumn{3}{c}{O32 \citep{kewley19_araa} -- Figure~\ref{fig:qgrad}}  & 
\colhead{} & 
\multicolumn{3}{c}{S2 \citep{kewley19_apj} -- Figure~\ref{fig:presgrad}}
\\
\cline{2-4} \cline{6-8} \cline{10-12}\noalign{\smallskip} 
\colhead{Galaxy} & 
\colhead{Central value} & 
\colhead{Gradient} & 
\colhead{Gradient} &
\colhead{} & 
\colhead{Central value}  & 
\colhead{Gradient} & 
\colhead{Gradient} &
\colhead{} & 
\colhead{Central value}  & 
\colhead{Gradient} & 
\colhead{Gradient} 
\\
\colhead{} & 
\colhead{(log(O/H)+12)} & 
\colhead{(dex/kpc)} & 
\colhead{(dex/R$_{25}$)} &
\colhead{} & 
\colhead{(log $U$)} & 
\colhead{(dex/kpc)} & 
\colhead{(dex/R$_{25}$)} &
\colhead{} & 
\colhead{(log P/k)} & 
\colhead{(dex/kpc)} & 
\colhead{(dex/R$_{25}$)} 
\\
}
\startdata 
NGC 1566	& $9.10\pm0.02$	& $-0.025\pm0.002$		& $-0.53\pm0.04$		&& $-3.19\pm0.04$ & $-0.010\pm0.003$ & $-0.21\pm0.06$	&& $5.65\pm0.10$ & $0.044\pm0.009$ & $0.94\pm0.19$ \\
NGC 2835	& $8.99\pm0.04$	& $-0.035\pm0.006$		& $-0.40\pm0.07$		&& $-3.04\pm0.05$ & $0.012\pm0.008$ & $0.14\pm0.09$		&& $5.9\pm0.3$ & $0.05\pm0.04$ & $0.6\pm0.5$  \\
NGC 3521	& $9.09\pm0.03$	& $-0.021\pm0.004$ 	& $-0.16\pm0.03$		&& $-3.07\pm0.06$ & $-0.003\pm0.007$ & $-0.023\pm0.05$	&& $5.70\pm0.15$ & $-0.03\pm0.02$ & $0.24pm0.16$ \\
NGC 5068	& $8.93\pm0.04$	& $-0.056\pm0.015$ 	& $-0.30\pm0.08$		&& $-3.13\pm0.07$ & $0.016\pm0.023$ & $0.09\pm0.12$		&& $5.4\pm0.5$ & $0.10\pm0.12$ & $0.5\pm0.8$  \\
NGC 5236	& $9.117\pm0.013$& $-0.006\pm0.003$	& $-0.028\pm0.013$	&& $-2.91\pm0.06$ & $-0.020\pm0.013$ & $-0.09\pm0.06$	&& $5.7\pm0.2$ & $-0.08\pm0.05$ & $-0.4\pm0.2$ \\
NGC 7793	& $8.85\pm0.05$	& $-0.040\pm0.017$ 	& $-0.19\pm0.08$		&& $-3.18\pm0.06$ & $-0.01\pm0.02$ & $-0.05\pm0.10$		&& $5.7\pm0.2$ & $-0.07\pm0.08$ & $-0.3\pm0.4$  \\
\enddata
\end{deluxetable*} 
\end{rotatetable*}

\section{Results}\label{sec:results}

\subsection{ISM Radial Gradients}\label{sec:gradients}

\subsubsection{Gas-Phase Oxygen Abundance Gradients}\label{sec:metgradients}
\begin{figure*}
\includegraphics[width=\linewidth]{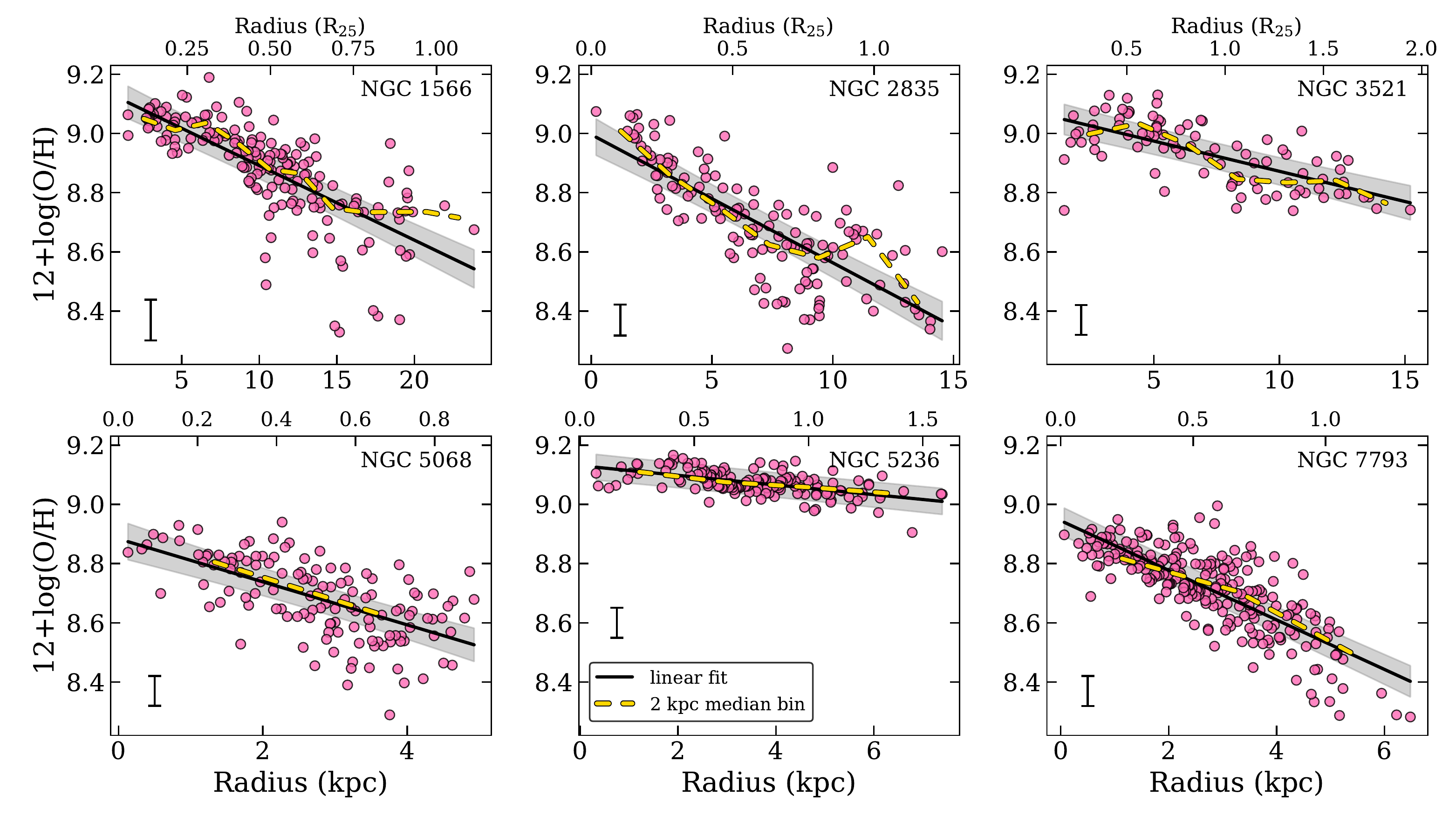}
\caption{Metallicity gradient profiles for each galaxy using the \n2o2 metallicity diagnostic \citep{kewley19_araa} with radius in kpc on the bottom axis and R$_{\rm 25}$ on the top axis. Overplotted are linear fits to the radial gradient (solid black), with the uncertainty in the fit shown as a gray region, and the median calculated for 2~kpc wide radial bins (dotted yellow). Uncertainties in the metallicity from propagating line flux errors (representative uncertainty is shown in the lower left corner of each plot) include the systematic uncertainties in the metallicity calibration and reddening correction. All galaxies except NGC 5236 show a significant negative gradient, where metallicities are higher in the central region relative to outer regions. }
\label{fig:metgrad_n202}
\end{figure*}

We present radial profiles of the gas-phase oxygen abundance in Figure~\ref{fig:metgrad_n202}. The geometric centers of the \HII\ regions are deprojected to the galactic disk frame using the inclination and positions angles as listed in (Table~\ref{tab:1}) under the assumption of a circular thin disk. We perform a linear fit to the radial gradient that includes the errors for all line flux measurements. We also show a rolling 2~kpc median bin to emphasize and highlight deviations of the \HII\ region metallicity from the single linear fit. The running median shows good agreement with the linear fit for NGC~3521, NGC~5068, and NGC~7793, indicating that the gradients are well-described with a single fit in these systems. NGC~1566 is an intriguing case, suggesting that the metallicity gradient starts to gradually flattens at galactocentric distance greater than $\sim$15~kpc. NGC~2835 also hints at evidence of a flattening in the metallicity profile at distances greater than $\sim$10~kpc, albeit with significant scatter. Flat abundance gradients appear to be fairly common features in the outer disks of star-forming galaxies \citep[e.g.,][]{sanchez-menguiano16a}. The metallicity plateau may be predominantly tracing the old stellar population in the outer disk and potentially is informative on radial migration and mixing scenarios within the ISM \citep{worthey05, vlajic09, bresolin09, minchev11, minchev12, hemler21}. 

The observed flattening of the gas-phase oxygen abundance may also reflect the Nitrogen to Oxygen abundance ratio N/O. This is a result of oxygen ($\alpha$-elements) primarily being produced through core-collapse supernovae (CCSNe) in massive stars ($>$8~M$_{\odot}$) whereas nitrogen is produced primarily by intermediate mass asymptotic giant branch (AGB) stars ($\sim$1--7~M$_{\odot}$) \citep{henry99, kobayashi11, kobayashi20}. Observations of the N/O as a function of metallicity in extragalactic \HII\ regions shows a bimodal behaviour, where the N/O ratio is approximately constant below 12 + log(O/H) $\sim$ 8.0 and increases at higher metallicities \citep[e.g., ][]{nicholls17}. This nebular scaling of N/O is interpreted through a combination of a primary and secondary production of nitrogen. The primary production of nitrogen gives a constant fraction with oxygen, enriched by CCSNe and predominately arises at low metallicity. The secondary production of nitrogen increases with the oxygen abundance and is enriched by intermediate-mass stars as they evolve, dominating at high O/H \citep{vila-costas93}. The outer disks of spiral galaxies have oxygen abundances that are typically below the level at which secondary nitrogen production starts to dominate. As a result of the primary production of nitrogen dominating within outer \HII\ regions, the radial trend of the N/O abundance ratio is often observed to flatten in the outer regions of disks beyond $\gtrsim0.75$~R$_{25}$ \citep[e.g., ][]{berg12, lopez-sanchez15, croxall16, rogers21}. 

The extremely shallow metallicity gradient for the \HII\ regions in NGC~5236 is a noticeable exception compared with the rest of the sample in Figure~\ref{fig:metgrad_n202}. The scatter around the single linear fit is correspondingly also very low, implying efficient mixing of the ISM across the disk of NGC~5236. \citet{bresolin09} also report a flat trend in the metallicity gradient and found that a low but persistent level of star formation over the past $\sim$3~Gyr is more than sufficient to lead to a substantial chemical enrichment within the outer disk of NGC~5236. The flat chemical abundance distribution across the extended disk of NGC~5236 could also be the result from interactions with dwarf galaxy members in the M83 Group \citep[e.g.,][]{vandenbergh80, thim03, silva-villa12}. We note that the neutral gas HI profile from the THINGS survey \citep{walter08, bigiel10} is flat and constant across the galaxy with an observed inhomogeneous filamentary distribution of HI in the outer disk at galactocentric distances $>0.6 R_{25}$ for regions in the galaxy with a mass column density larger than 0.5 M$_{\odot}$~pc$^{-2}$; this warped structure in the outer region is consistent with a possible signature of past galaxy encounters \citep{rogstad74}. 

All the galaxies in our survey report prior metallicity gradients calculated from a broad range of nebular strong-line and T$e$-based diagnostics used throughout the literature (Section~\ref{sec:galaxies}). Metallicities derived using theoretical compared to empirical calibrations produce significantly different abundance estimations \citep[e.g.,][]{kewley08, lopez-sanchez12b}. Thus, oxygen abundances that are obtained in different studies using different calibrations can be significantly different by up to 1~dex. Directly comparing abundances from different studies directly to each other cannot be done as abundances from between studies are not homogeneous. \citet{pilyugin14} determine all the abundances in a uniform manner for 130 galaxies, which includes all galaxies in our survey with the exception of NGC~1566. The uniform gas-phase metallicity gradients as reported in \citet[][listed in Section~\ref{sec:galaxies}]{pilyugin14} agree with our measured metallicity gradients within the errors, listed in Table~\ref{tab:gradients_hii}.

\subsubsection{Ionization Parameter Gradients}\label{sec:qgradients}
\begin{figure*}
\includegraphics[width=\linewidth]{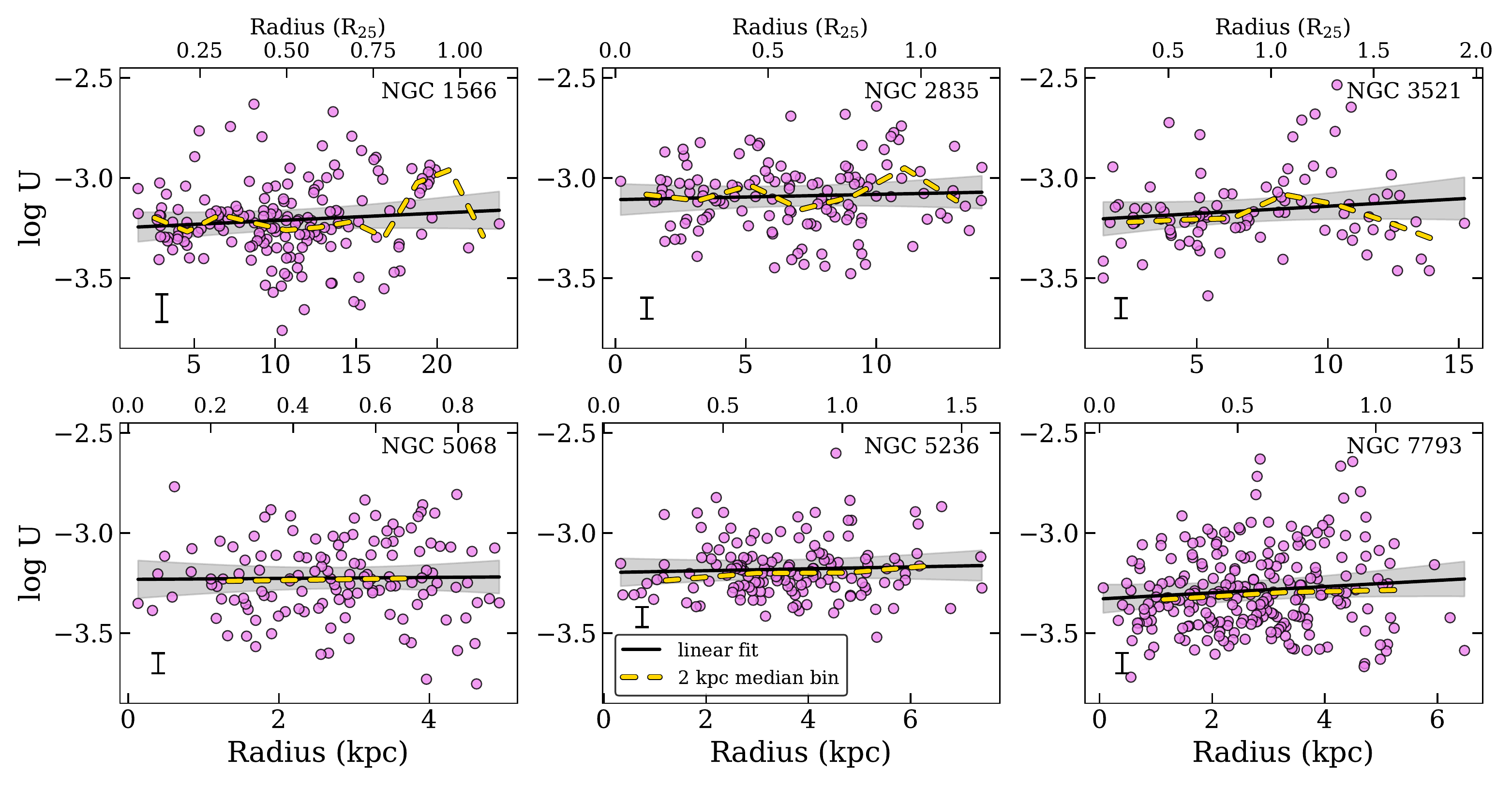}
\caption{Ionization parameter $U$ gradient profiles for each galaxy using the O32 diagnostic \citep{kewley19_araa} with radius in kpc on the bottom axis and R$_{\rm 25}$ on the top axis. Overplotted are linear fits to the radial gradient (solid black), with the uncertainty in the fit shown as a gray region, and the median calculated for 2~kpc wide radial bins (dotted yellow).  Representative uncertainty is shown in the lower left corner of each plot. No trends are evident with ionization parameter $U$ and galaxy radius.}
\label{fig:qgrad}
\end{figure*}

Figure~\ref{fig:qgrad} presents the radial profiles of the ionization parameter $U$ using the [OIII]/[OII] emission line ratios. The linear fit to the radial gradient includes the errors for all line flux measurements. We also show a rolling 2~kpc median bin, which shows good agreement with the linear fit. All ionization parameter radial gradients are either flat or negligibly positive, a result also found in prior studies \citep{poetrodjojo18, kreckel19}. This suggests that there is a very narrow range of ionization parameter and pressure values and that there is no dependence on galactocentric distance within the galaxy. Indeed, observations support that ionization parameter values are remarkably uniform across normal star-forming galaxies \citep{kewley19_araa}. We do note, however, that strong emission lines in the ultraviolet, optical, and infrared spectral regions trace different ranges of ionization energies and zones within a nebula. Thus, the [OIII]/[OII] emission line ratios we use to calculate the ionization parameter $q$ may only trace a narrow range in ionization parameter and using a different ionization diagnostics probing other energy ranges may show different results. 

The coefficients for the radial fit to the ionization parameter gradients are listed in Table~\ref{tab:gradients_hii}.

\subsubsection{Pressure Gradients}\label{sec:presgradients}
\begin{figure*}
\includegraphics[width=\linewidth]{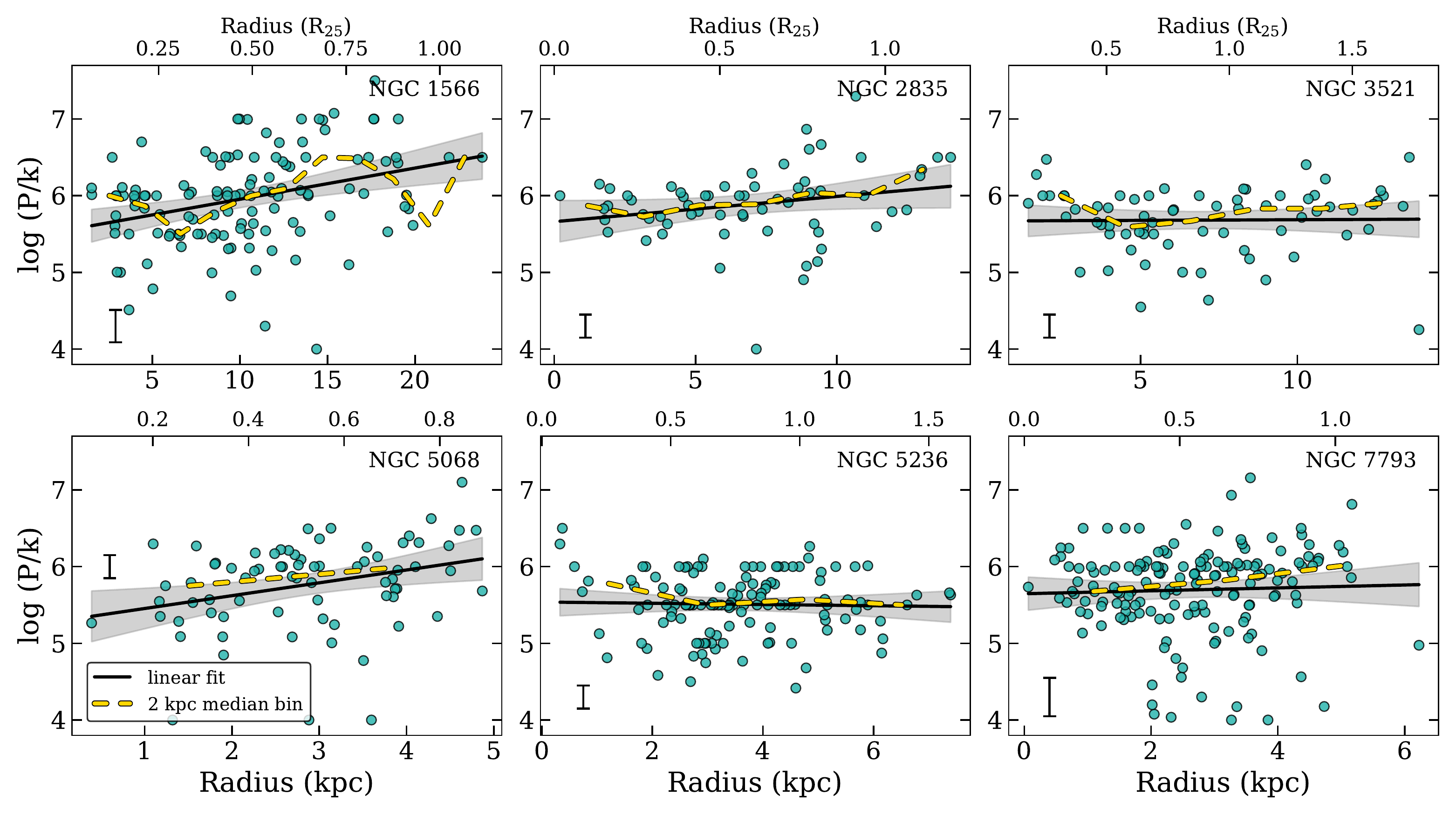}
\caption{Pressure log ($P/k$) gradient profiles for each galaxy using the [SII]$\lambda 6717$/[SII]$\lambda 6730$ doublet line ratios diagnostic \citep{kewley19_apj}. Radius in kpc is shown on the bottom axis and R$_{\rm 25}$ displayed on the top axis. Overplotted are linear fits to the radial gradient (solid black), with the uncertainty in the fit shown as a gray region, and the median calculated for 2~kpc wide radial bins (dotted yellow). Representative uncertainty is shown in the lower left corner of each plot. All galaxies show pressure radial gradients that are either flat or negligibly positive.} 
\label{fig:presgrad}
\end{figure*}

Figure~\ref{fig:presgrad} presents the radial profiles of the ISM electron density pressure using the [SII]$\lambda$6717/[SII]$\lambda$6730 emission line ratios. The linear fit to the radial gradient includes the errors for all line flux measurements. We also show a rolling 2~kpc median bin, which shows good agreement with the linear fit. Similar to ionization parameter $U$ (Section~\ref{sec:qgradients}), all pressure radial gradients are either flat or negligibly positive, albeit with significant scatter.

The coefficients for the radial fit to the pressure gradients are listed in Table~\ref{tab:gradients_hii}.

\subsection{Azimuthal Variations}\label{sec:azimuthal}
To account for the metallicity offsets between galaxies introduced by the mass--metallicity relation \citep{tremonti04} and discrepancies of up to 1 dex in 12+log(O/H) arising from the adoption of different metallicity calibrations \citep{kewley08}, we subtract off the fitted radial metallicity gradients, and consider the residuals in the metallicity from the linear trend $\Delta$(O/H). We do not subtract off the linear radial measurement to the ionization parameter and pressure radial profiles as they already display flat gradients in their radial distributions. The change in the metallicity $\Delta$(O/H) will aid in the inspection and analysis of azimuthal effects in addition to galactic features like bars and spiral arms on the metallicity, ionization parameter, and pressure distribution. 

The spatial distribution of oxygen abundance in the ISM is well-characterized by a single linear gradient (Figure~\ref{fig:metgrad_n202}). Deviations from the linear gradient in the azimuthal direction can provide information on how \HII\ region oxygen abundance is mixed within the ISM when the stars and ISM gas orbit in the galactic potential. In general, we find a lack of correlation between the metallicity residuals and the spiral patterns. There is one exception, NGC~5236, where the metallicity residuals correlate strongly with the spiral pattern and exhibit an enhanced metallicity along the spiral arms in this system. 

The azimuthal variation of oxygen is indicative of gas that undergoes localized, sub-kiloparsec-scale self-enrichment that experiences efficient mixing-induced dilution \citep{kennicutt96, sanchez-menguiano16, vogt17, ho17, ho18, ho19}. This effect of dynamical local enrichment of oxygen enhancement modulated by a spiral-driven, mixing and dilution or the interaction with dwarf galaxies in the last $\sim1$~Gyr \citep[e.g., ][]{thim03} are likely driving the observed azimuthal metallicity distributions in NGC~5236. 

We do not observe noticeable azimuthal variations in the sample that are cleanly associated with the spiral pattern of a galaxy. Instead, we find that \HII\ regions with enhanced or reduced metallicity are uniformly spatially distributed across the full disk of all the galaxies. Thus, we conclude that while a spiral pattern plays a role in organizing the ISM, the spiral pattern alone cannot not establish the azimuthal variations (or lack therefore) that we observe. Instead, metal abundance variations are more likely to be driven by correlations with the local physical conditions that drive localized enrichment across galaxy disks (see Section~\ref{sec:delOH}).

\begin{figure*}
\includegraphics[width=0.99\linewidth]{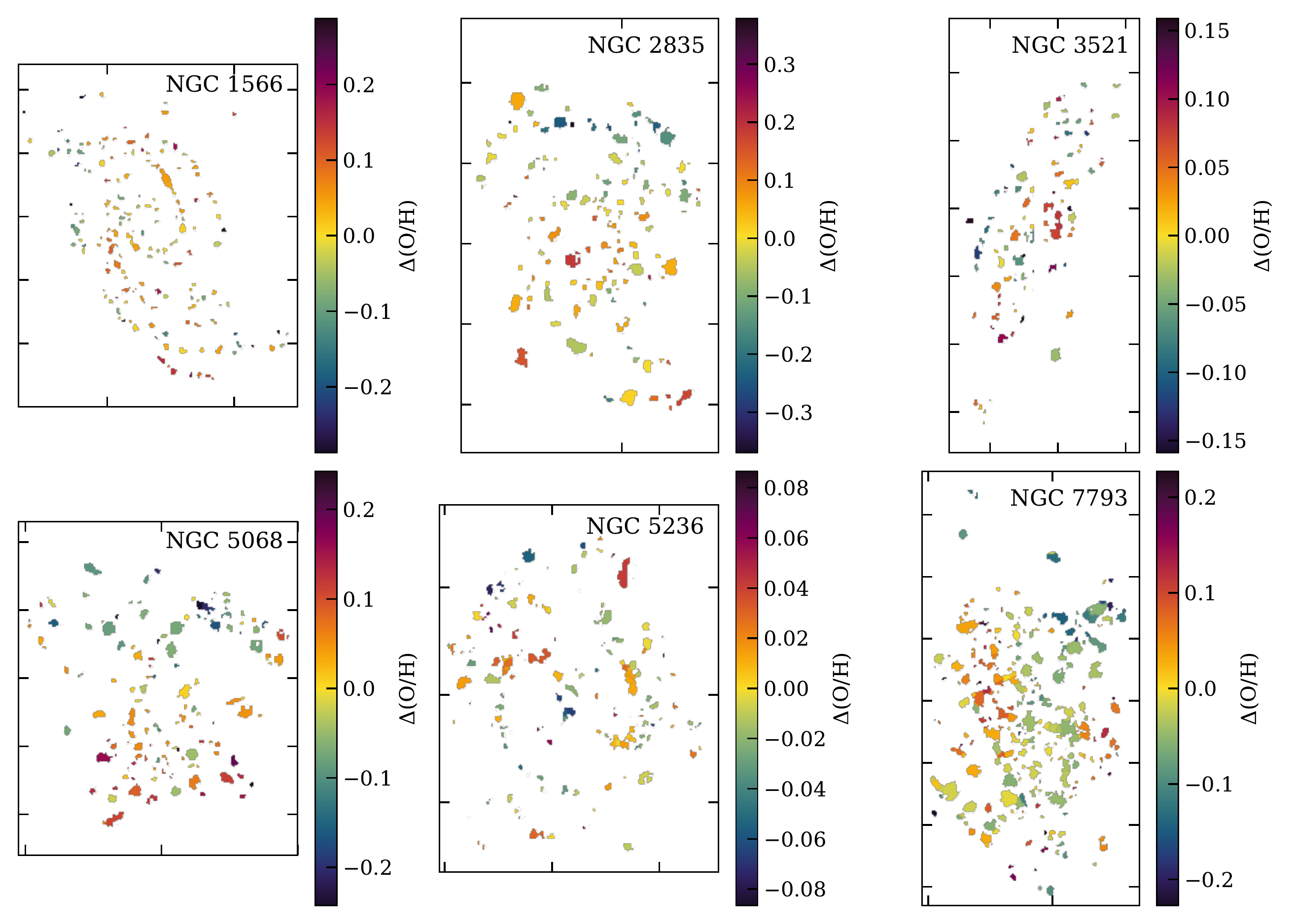}
\caption{Oxygen abundance map after subtracting off the best-fit radial abundance gradient from Figure~\ref{fig:metgrad_n202}. Due to a lack of strong spiral galaxies in our survey, the azimuthal variations of oxygen abundances is not significant in any galaxy with the exception of NGC~5236 with an enhancement of oxygen abundance along the arms (see Section~\ref{sec:azimuthal}). }
\label{fig:delOH}
\end{figure*}

As the ionization parameter $U$ and pressure log(P/$k$) in the \HII\ regions are demonstrated to be remarkably uniform with no radial trend, we do not calculate the residual ionization parameter or pressure values by subtracting the best-fit radial gradient fits.

\subsection{Impact of Local Physical Conditions on the Properties of \HII\ Regions}\label{sec:delOH}
In order to constrain the local physical conditions and possible linkage to the localized enrichment of the ISM, we investigate the relationship between the metallicity and residual metallicity $\Delta$(O/H) as a function of the ionization parameter and ISM pressure of the \HII\ regions in each individual galaxy. 

\begin{figure*}
\includegraphics[width=0.495\linewidth]{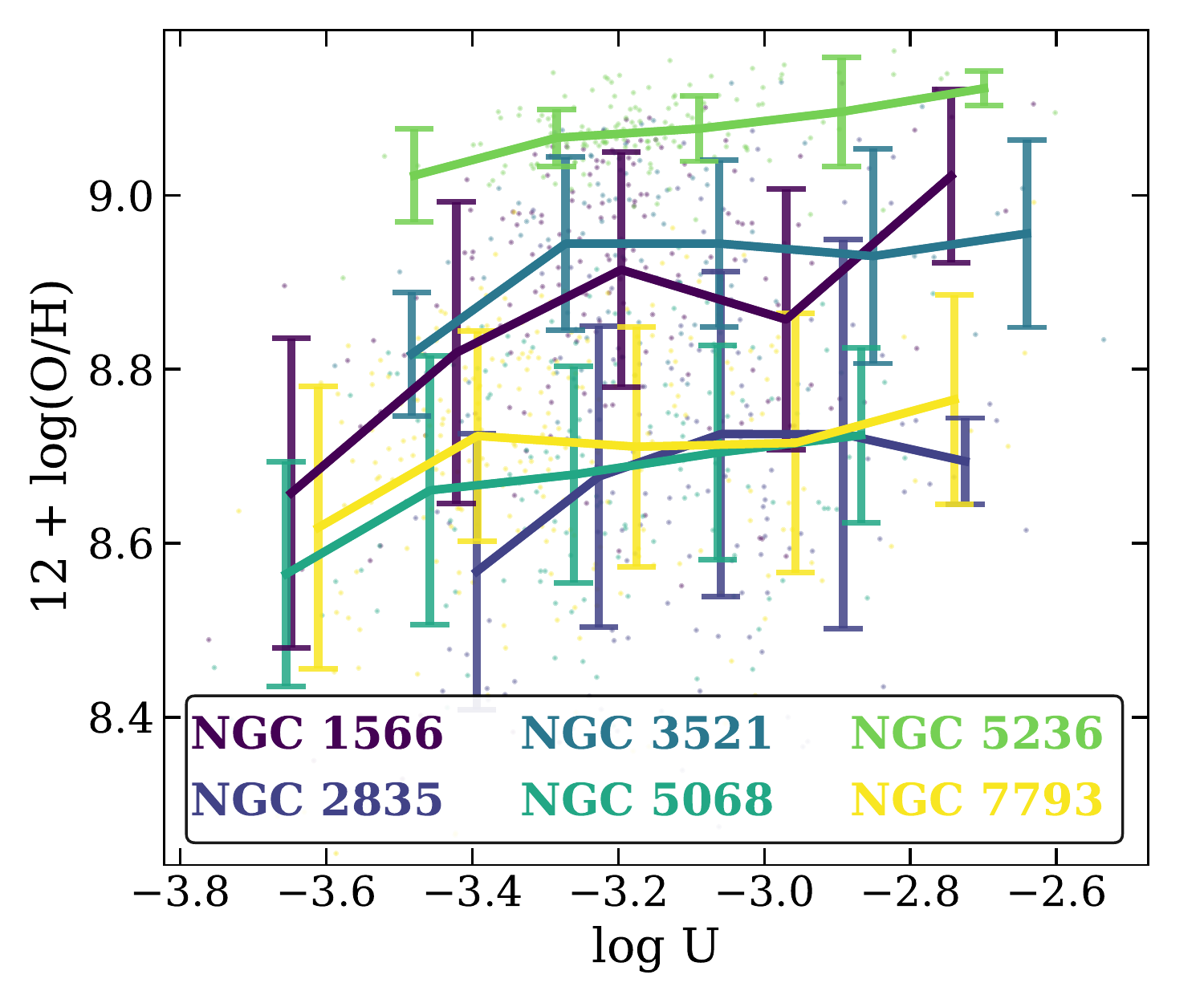}
\includegraphics[width=0.495\linewidth]{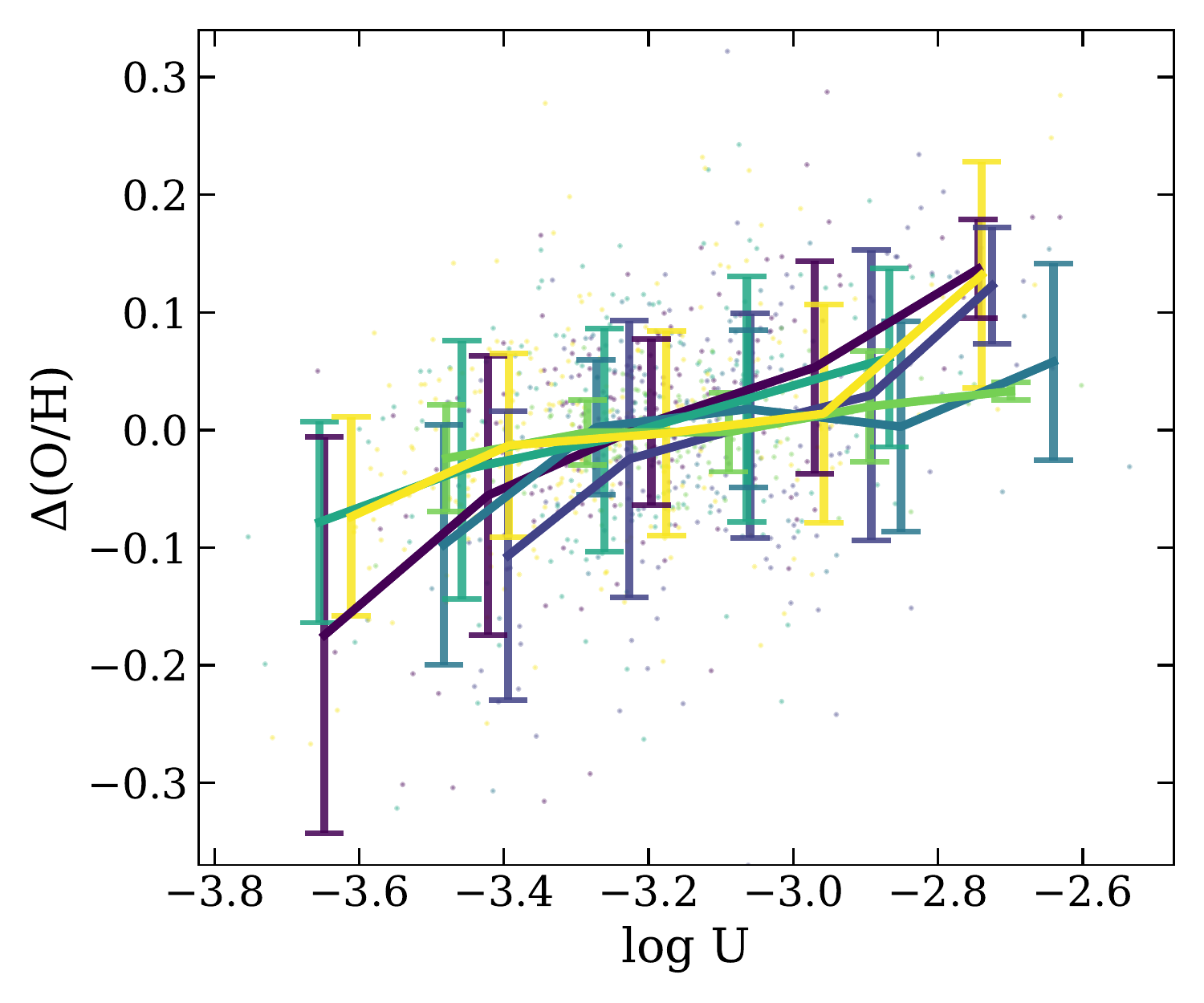}
\caption{\emph{Left:} 12+log(O/H) metallicity of each \HII\ region versus the ionization parameter $U$ colored by each individual galaxy. The error bars represent the 1$\sigma$ scatter for each galaxy. All galaxies show a positive correlation, with the vertical offset in metallicity between the galaxies determined by the stellar mass-metallicity relation. \emph{Right:} The metallicity offset from the linear radial gradient $\Delta$(O/H) versus the ionization parameter $U$. There is a significant positive relationship between the residual metallicity $\Delta$(O/H) and the ionization parameter $U$ in the \HII\ regions in each galaxy. Table~\ref{tab:4} lists the Spearman rank correlation coefficients and the corresponding p-values for the statistical significance.}
\label{fig:z_vs_q}
\end{figure*}

In Figure~\ref{fig:z_vs_q}, we plot the metallicity as a function of the ionization parameter. The vertical offset between different galaxies is simply set by the difference in stellar mass of the systems (i.e., the mass-metallicity relation). For all galaxies, we observe a positive correlation between the metallicity and ionization parameter of the \HII\ regions. Because the ionization parameter (Figure~\ref{fig:qgrad}) shows no radial gradient, any correlation with metallicity is not a result of correlated observables. In this work, we have carefully selected a metallicity diagnostic that no dependence on ionization parameter and only marginal dependence on the ISM pressure for $4 \leq \log(P/k) \leq 8$ \citep{kewley19_araa}. In addition, the [NII]/[OII] emission line ratio is also the least sensitive optical diagnostic to the presence of an AGN or DIG contamination \citep{zhang17}. Figure~\ref{fig:z_vs_q} also shows the relationship between residual metallicity $\Delta$(O/H) and ionization parameter, in order to account for and remove the the metallicity offsets between galaxies introduced by the mass-metallicity relation. Table~\ref{tab:4} lists the Spearman's rank correlation coefficient $\rho$ for each galaxy and the corresponding p-value. All galaxies have high statistical significance between metallicity and ionization parameter ($\rho<$ 0.05) with a tighter correlation observed between the residual metallicity $\Delta$(O/H) ($\rho<$ 0.001). 

Prior studies have found mixed results regarding a correlation between the ionization parameter and metallicity, where some authors report a negative correlation \citep{bresolin99}, lack of a significant trend \citep{kennicutt96, poetrodjojo18}, or a positive correlation \citep{dopita14, kreckel19}. The positive and significant correlation we recover in all of our galaxies between the relative enhancement in the \HII\ region metallicity $\Delta$(O/H) and the ionization parameter indicates a very strong relation that links the local physical conditions of the gas-phase ISM to the localized enrichment of the gas-phase ISM. It is unsurprising that prior work may have resulted in a lack of a trend compared to the strong correlation observed in this study because prior studies with global metallicities may have washed out the correlation we identify arising within individual galaxies. \citet{kreckel19} also report a positive correlation between the relative enhancement in the \HII\ region metallicity $\Delta$(O/H) and the ionization parameter in nearby disk galaxies with MUSE observations. They find small systematic scatter in $\Delta$(O/H) where \HII\ regions with enhanced abundances show offsets of $\sim$0.05~dex whereas our data encompass a broader range in the scatter with $\Delta$(O/H) offsets of $\sim$0.1~dex. Despite this, we recover a remarkably similar trend as observed by \citet{kreckel19}. 

\citet{dopita14} present a scenario where that the observed positive correlation between metallicity and ionization parameter of \HII\ regions is attributed to the correlation between the star formation rate surface density and ionization parameter underpinned by a geometrical effect. In this scenario, the densest star-forming regions contain \HII\ regions that are embedded co-spatially with molecular clouds. In these dense star-forming regions, the molecular clouds undergo radiation pressure-dominated photo-ablation due to their co-mixture with young, massive stars that are exciting both the ionized gas and the molecular gas under extremely high ISM pressures of P/k $\sim$ 10$^7$ cm$^{-3}$~K \citep{smith06, westmoquette07}. Such a geometric effect could be driving the observed positive correlation between metal abundance enrichment and ionization parameter as the active star-forming regions have a different distribution of molecular gas which in turn favors a higher ionization parameter. 

We highlight that we are studying normal star-forming galaxies and not active star-forming regions with distributions of molecular gas that favor higher ionization parameters, as was done in \citet{dopita14}. This is further supported by the fact that more metal-rich \HII\ regions have no strong systemic trend with galaxy environment and as such are not preferentially located along the spiral arm ridge but instead, are located throughout the star-forming disk (Figure~\ref{fig:delOH}). We advocate for a different explanation as the ISM conditions of the \HII\ regions in our normal star-forming disk galaxies are not consistent with the physical conditions of LIRGs presented in \citet{dopita14}. 

The positive trend between metallicity offset and ionization parameter may reflect a star formation history that has locally enriched the ISM from the most recent generation of stars. 
The oxygen yield per stellar generation is, however, quite uncertain, and the uncertainty introduced by assuming different sets of stellar yields can be quantified by changes in log(O/H) of $\sim0.02-0.2$ with uncertainties of $\sim0.2$~dex \citep[e.g., ][]{woosley95, pilyugin07, kudritzki15, vincenzo16} and strong dependencies on the upper mass cutoff of the IMF. We measure systematic scatter in $\Delta$(O/H) with offsets of $\sim$0.1~dex across all of our galaxies. While the level of enrichment across the \HII\ regions in our sample of galaxies is consistent with self-enrichment from a single burst of star formation, we cannot rule out continuous star formation that may be occurring across larger star-forming complexes than these \HII\ regions are embedded in.

\begin{figure*}
\includegraphics[width=0.495\linewidth]{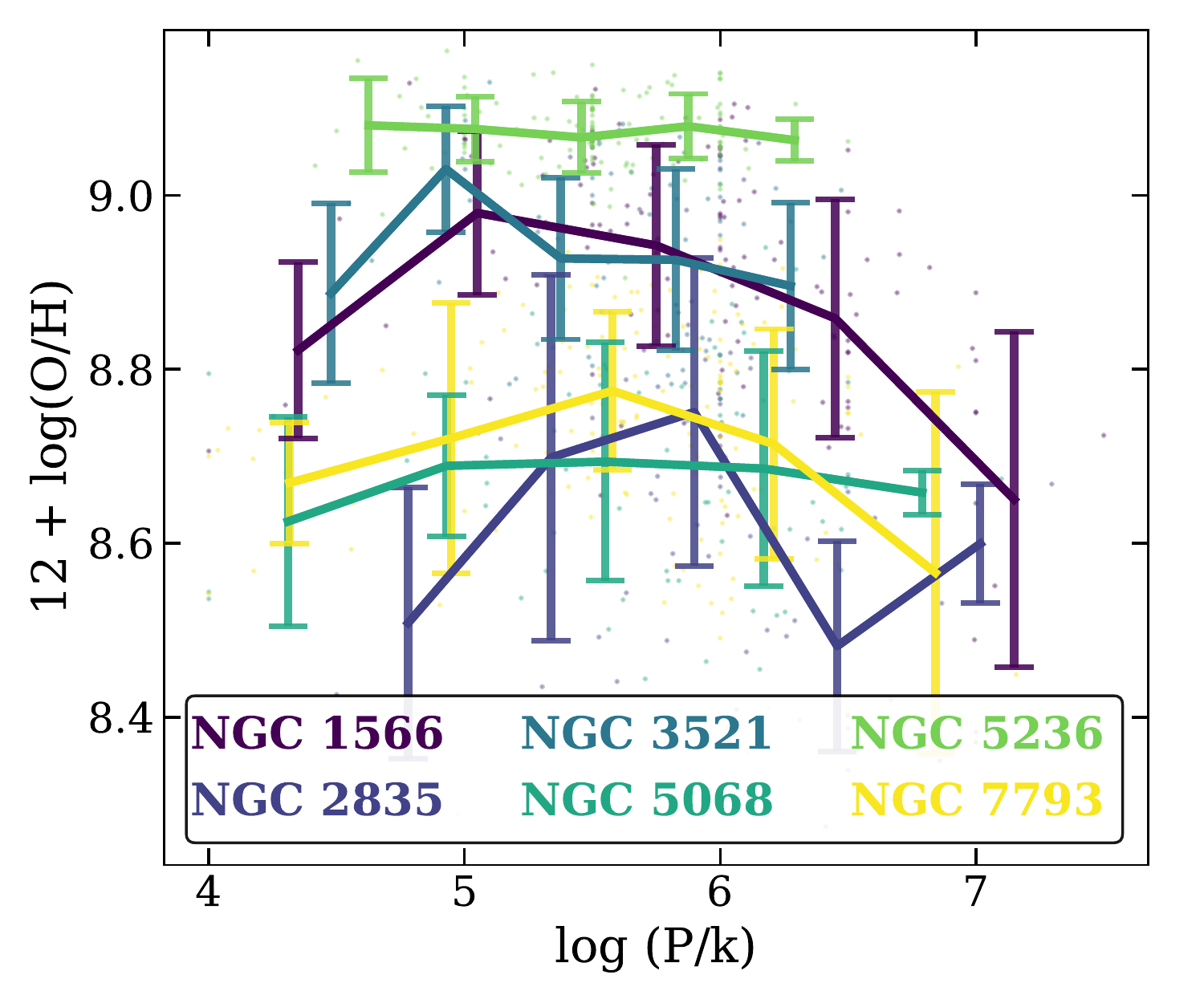}
\includegraphics[width=0.495\linewidth]{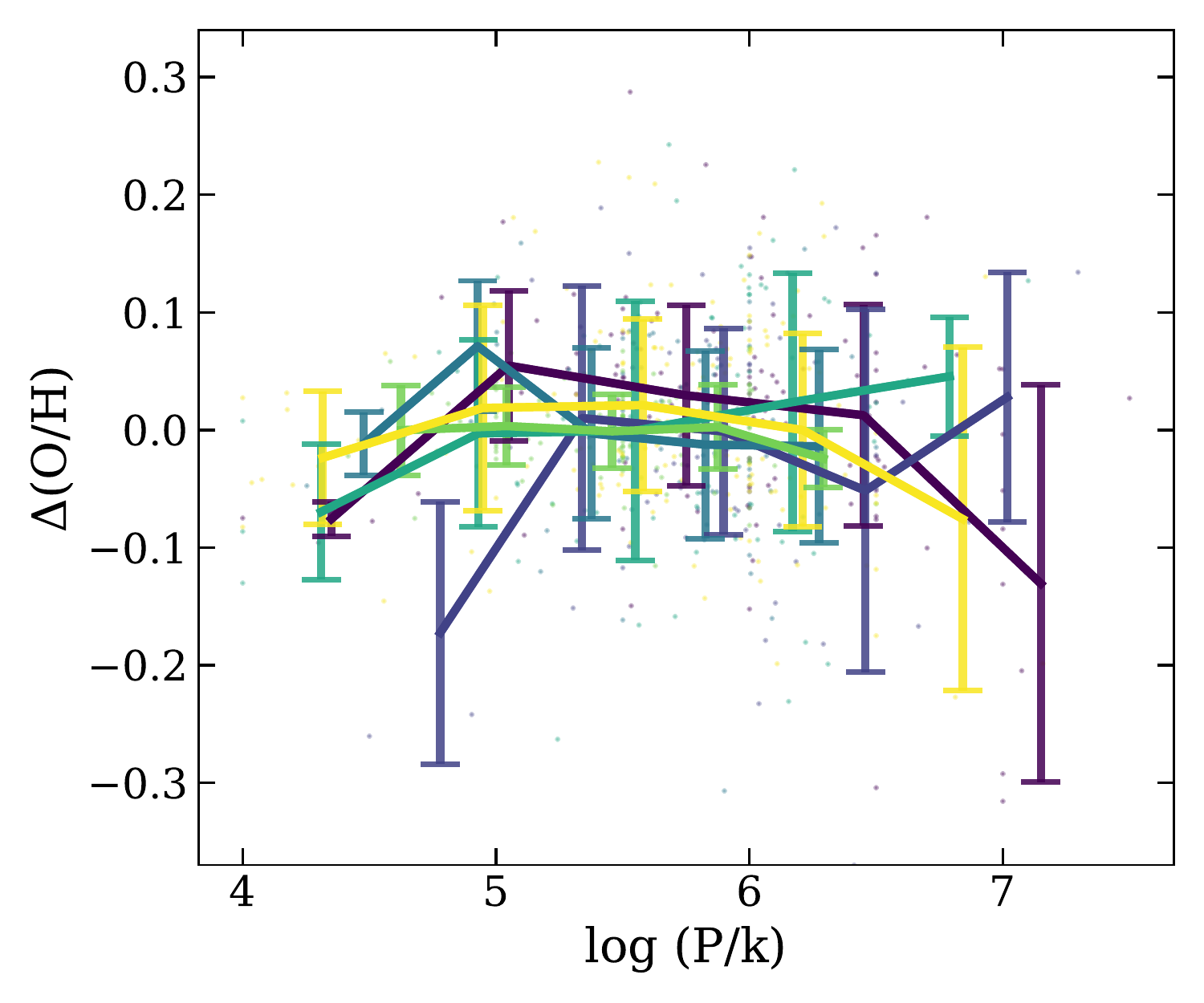}
\caption{Same as Figure~\ref{fig:z_vs_q} but for the 12+log(O/H) metallicity (\emph{left}) and residual metallicity $\Delta$(O/H) (\emph{right}) versus the ISM pressure log(P/$k$). There is no trend observed between the residual metallicity $\Delta$(O/H) and the ISM pressure. Table~\ref{tab:4} lists the Spearman rank correlation coefficients and the corresponding p-values for the statistical significance.}
\label{fig:z_vs_pres}
\end{figure*}

When we look for a possible correlation between the localized enrichment of \HII\ regions with the ISM pressure as traced by the [SII] doublet, we find no significant correlation between the ISM pressure and metallicity or residual metallicity (Figure~\ref{fig:z_vs_pres}). Table~\ref{tab:4} lists the Spearman rank correlation coefficients and the corresponding p-values are listed for a hypothesis test whose null hypothesis is that two sets of data are uncorrelated.

Future work to connect the different phases of the ISM conditions inferred from the young stellar populations (rest-frame ultraviolet spectroscopy) and molecular gas (CO) will help create a direct link between the physics occurring in the different ISM phases. However, it is important to remember that different phases are not necessarily co-spatial with the ionized phase of \HII\ regions. The importance of (relatively inefficient) large scale mixing timescales and the ability for multiple generation of star formation to occur and locally enrich \HII\ regions before mixing has finished occurring to give rise to the uniformly distributed, locally enriched the ISM is further discussed in Section~\ref{sec:discussion_1}.

\subsection{Impact of Spatial Resolution on Observed Gradients}\label{sec:1kpc}
The TYPHOON observations in this study are well matched to the typical spatial scales of \HII\ regions of tens to a few hundreds of parsecs \citep{azimlu11}. This provides the enormous benefit of identifying \HII\ regions and reduces the bias of having multiple ionization sources contributing to a single resolution element in observations that lack the spatial resolution to resolve individual \HII\ regions. The effect of lower resolution dilutes the metallicity measurements, which flattens observed radial gradients. While the effect of a decrease in spatial resolution leading to a flattening of observed metallicity gradients has been well-demonstrated \citep{yuan13b, poetrodjojo19, acharyya20}, the effect of a flattening in ionization parameter or pressure gradients due to the decreasing spatial scales/smoothing has not been thoroughly investigated until now \citep[e.g.,][]{poetrodjojo18}.

To constrain the impact of lower resolution on our observations and to facilitate ease of comparison with surveys that have typical resolution scales of $\sim$1~kpc such as the Sydney-Australian-Astronomical-Observatory Multi-object Integral-Field Spectrograph survey \citep[SAMI;][]{croom21} and the Mapping Nearby Galaxies at APO survey \citep[MaNGA;][]{bundy15}, we re-bin the native resolution of each galaxy to a spatial scale of 1~kpc. 
At this scale, large morphological features such as the spiral arms are no longer distinguishable. The binning of the datacubes to lower resolution scales is done on the original datacube. Each rebinned datacube after downsampled to 1~kpc following the method by \citet{poetrodjojo19} where we sum the flux and add the errors in quadrature. We then re-processed the binned datacube and summed fluxes with LZIFU to extract the total line flux for the emission lines. 

\begin{figure*}
\includegraphics[width=\linewidth]{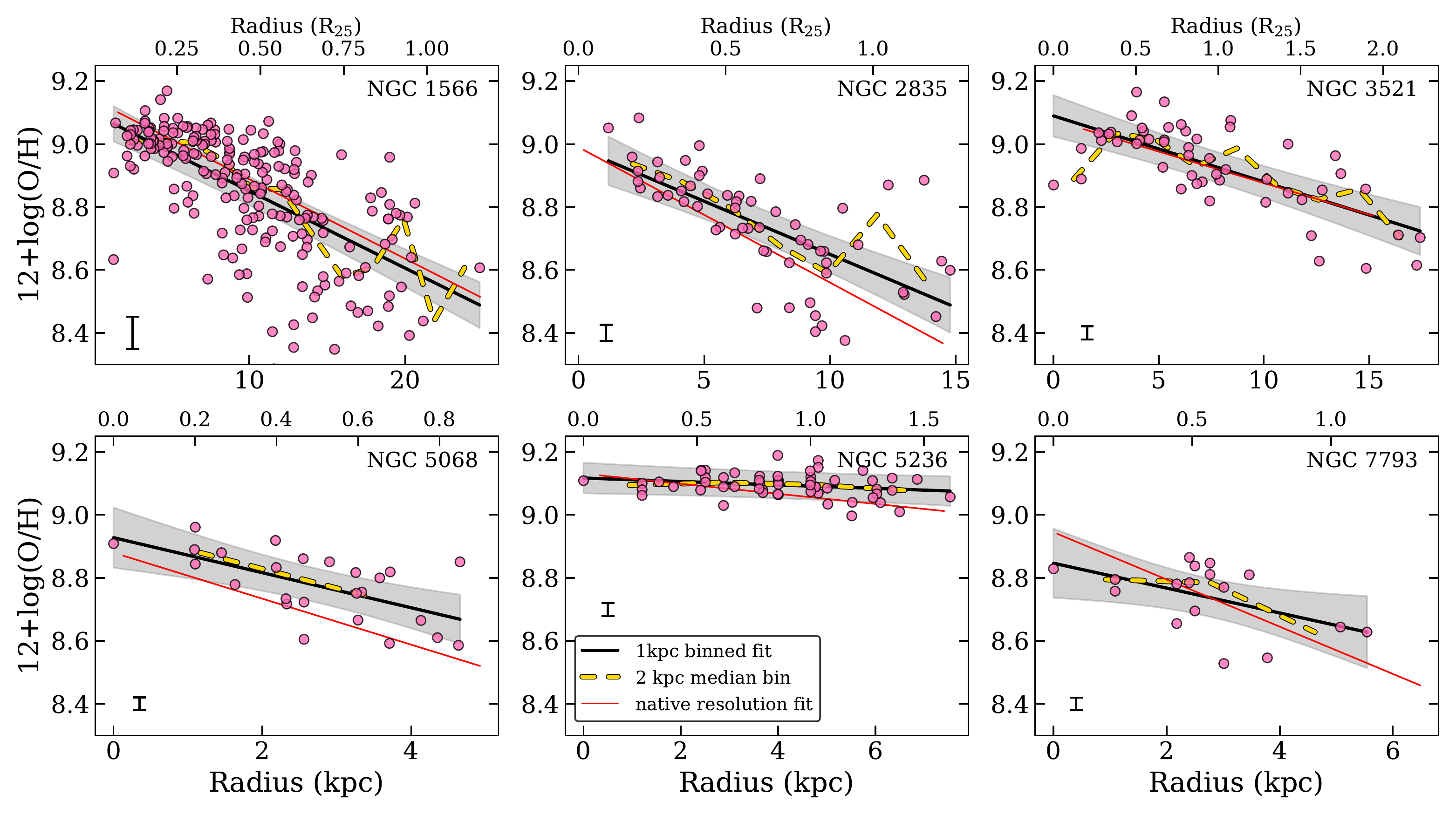}
\caption{Metallicity gradient profiles similar to Figure~\ref{fig:metgrad_n202} but for 1~kpc spaxels. Radius in kpc is shown on the bottom axis and R$_{\rm 25}$ displayed on the top axis. Overplotted are linear fits to the radial gradient (solid black), with the uncertainty in the fit shown as a gray region, the median calculated for 2~kpc wide radial bins (dotted yellow), and the linear fit to the native resolution \HII\ regions (red; Figure~\ref{fig:metgrad_n202}). Representative uncertainty is shown in the lower left corner of each plot. With the exception of NGC~1566 and NGC~3521, the rebinned data to simulate low-resolution observations at varying resolution scales show metallicity diagnostics that are more shallow compared to the native \HII\ region resolution (red line), likely a contribution of DIG contamination to the star-forming \HII\ region metallicity. } 
\label{fig:metgrad_n2o2_1kpc}
\end{figure*}

\begin{figure*}
\includegraphics[width=\linewidth]{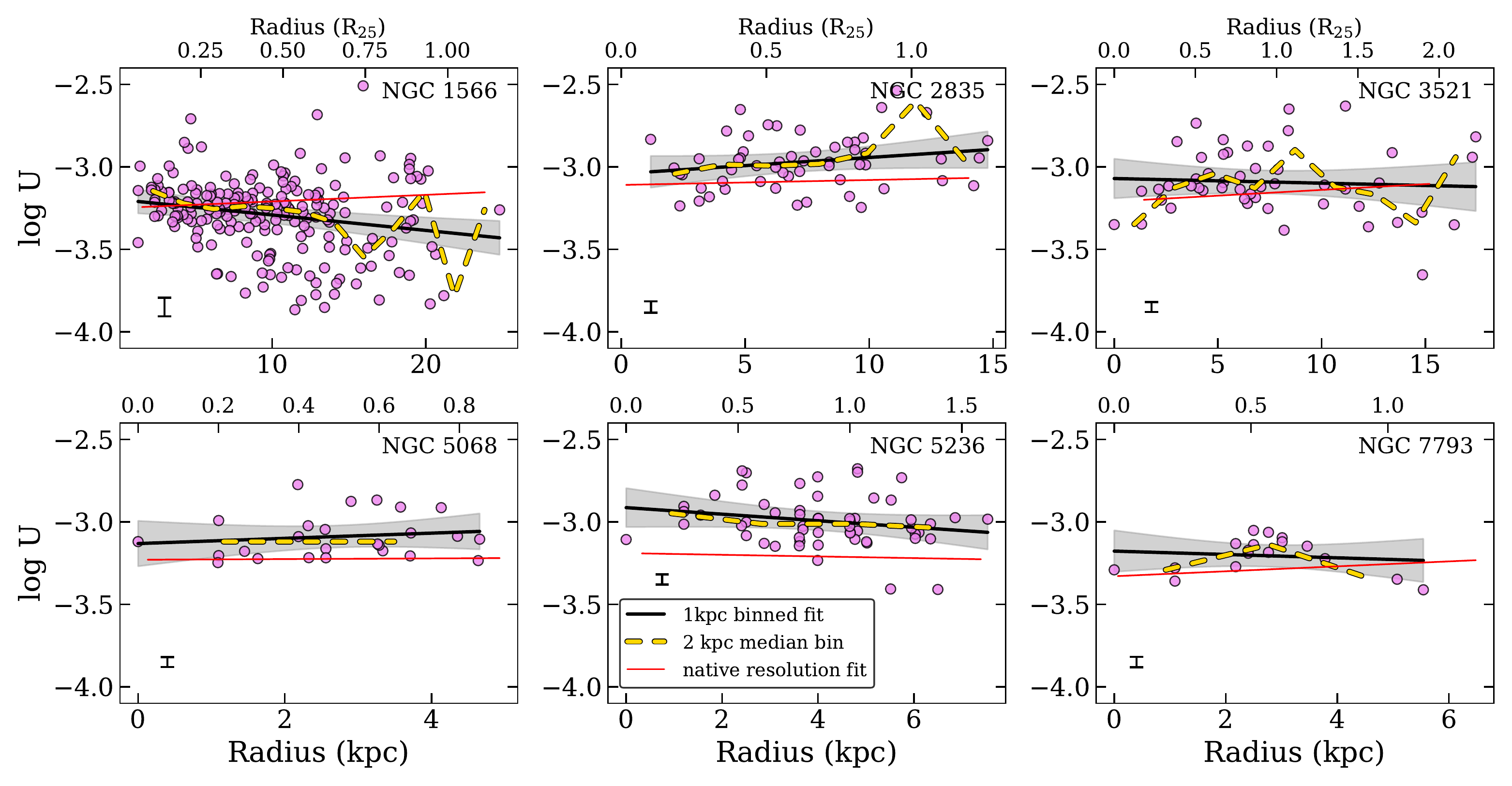}
\caption{Ionization gradient profiles similar to Figure~\ref{fig:qgrad} but for 1~kpc spaxels. Radius in kpc is shown on the bottom axis and R$_{\rm 25}$ displayed on the top axis. Overplotted are linear fits to the radial gradient (solid black), with the uncertainty in the fit shown as a gray region, the median calculated for 2~kpc wide radial bins (dotted yellow), and the linear fit to the native resolution \HII\ regions (red; Figure~\ref{fig:qgrad}). Representative uncertainty is shown in the lower left corner of each plot. We do not observe a flatting of the ionization parameter radial gradient for the rebinned 1~kpc spaxels. The absolute scaling of ionization parameter is slightly larger at lower resolution compared to the native \HII\ region resolution. }
\label{fig:qgrad_1kpc}
\end{figure*}

\begin{figure*}
\includegraphics[width=\linewidth]{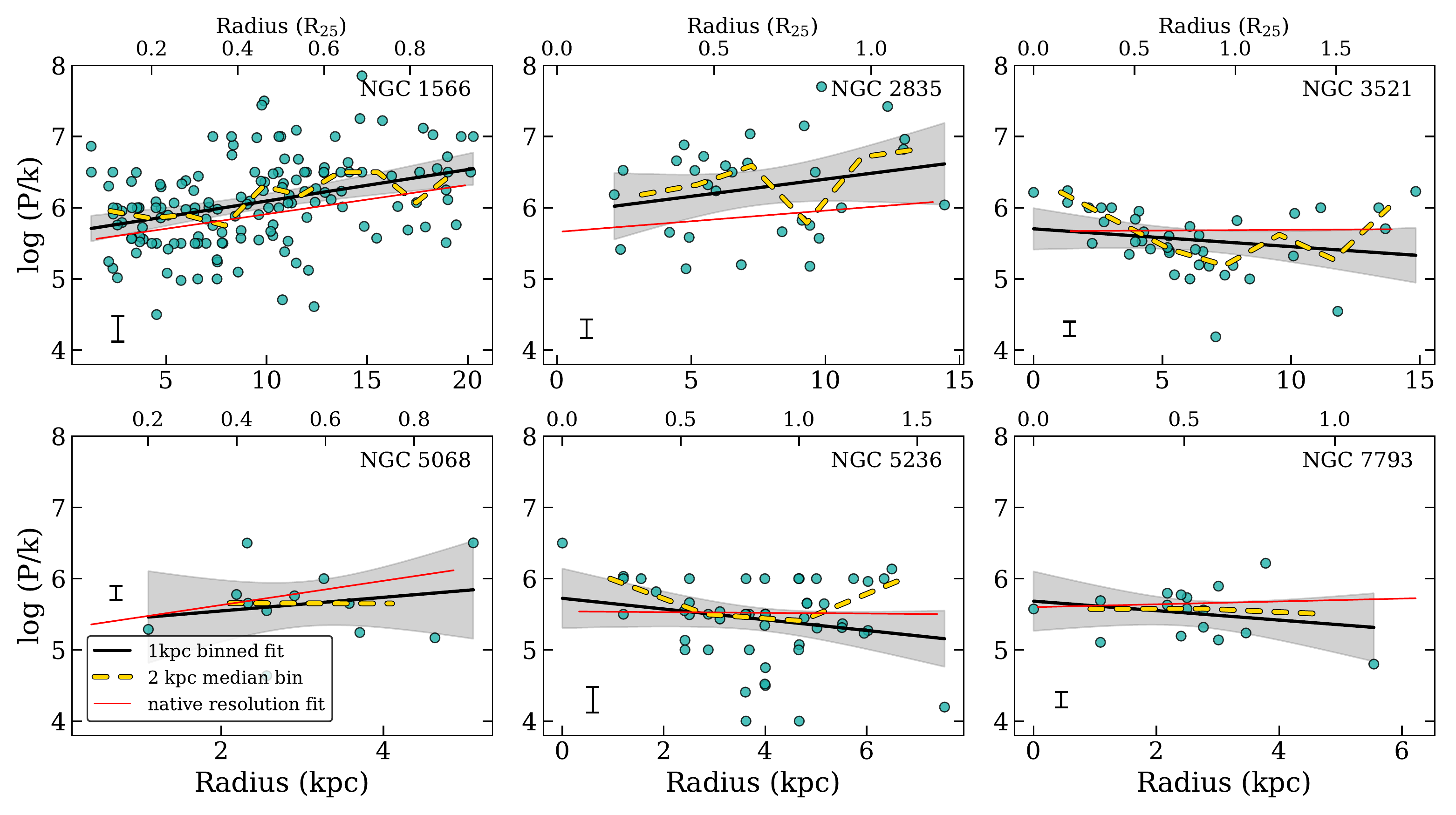}
\caption{Pressure gradient profiles similar to Figure~\ref{fig:presgrad} but for 1~kpc spaxels. Radius in kpc is shown on the bottom axis and R$_{\rm 25}$ displayed on the top axis. Overplotted are linear fits to the radial gradient (solid black), with the uncertainty in the fit shown as a gray region, the median calculated for 2~kpc wide radial bins (dotted yellow), and the linear fit to the native resolution \HII\ regions (red; Figure~\ref{fig:presgrad}). Representative uncertainty is shown in the lower left corner of each plot. We do not observe a flatting of the pressure radial gradient for the rebinned 1~kpc spaxels. In addition, the absolute scaling of pressure is marginally larger at lower resolution compared to the native \HII\ region resolution. }
\label{fig:presgrad_1kpc}
\end{figure*}

We re-calculate the ISM properties for the re-binned 1~kpc spaxel datacubes, presented in Figure~\ref{fig:metgrad_n2o2_1kpc} (metallicity), Figure~\ref{fig:qgrad_1kpc} (ionization parameter), and Figure~\ref{fig:presgrad_1kpc} (pressure). The linear fits to the rebinned 1~kpc gradient results are reported in Table~\ref{tab:gradient_1kpc}.

We can inspect Figures~\ref{fig:metgrad_n2o2_1kpc}, \ref{fig:qgrad_1kpc}, and \ref{fig:presgrad_1kpc} for the relative change in both the absolute value and the radial gradient between the native \HII\ region resolved dataset and the rebinned 1~kpc spaxels. For the 1~kpc metallicity radial profiles, the typical flattening in the gradient is observed. In general, the farthest galaxies are least affected by the loss in resolution, whereas the closest galaxies show gradients that are significantly flattened compared to their native resolution gradients. This is not unexpected, as the farthest galaxy in this study, NGC~1566, with resolution of 143~pc/pixel, is approaching spatial scales that are not sub-\HII\ region in size and the \HII\ regions in this galaxy may already be mixing multiple regions with different ionization sources and suffering DIG contamination, thus flattening the observed native gradient to begin with. 

Surprisingly, we do not observe a flatting of the ionization parameter or pressure radial gradients for the rebinned 1~kpc spaxels 
(Figure~\ref{fig:qgrad_1kpc} and Figure~\ref{fig:presgrad_1kpc}, respectively). For the lower resolution results, we find that the radial gradients either remain the same or slightly steepen. We point out, however, that we have a very limited number of spaxels for most of the rebinned 1~kpc galaxies. When this fact is combined with the flat radial gradients observed natively in both the ionization parameter and pressure, it suggests that resolution has little impact on the radial gradients of both ionization parameter and pressure of \HII\ regions. 

The offset of pressure and ionization parameter hint at a slight impact due to change in spatial resolution. As also shown in Figure~\ref{fig:qgrad_1kpc} and Figure~\ref{fig:presgrad_1kpc}, the absolute scaling of both pressure and ionization parameter are marginally larger at lower resolution compared to the native \HII\ region resolution, however for the ISM pressure, the offsets are consistent within the 1$\sigma$ spread. The metallicity offset is not impacted by resolution and remains relatively unchanged between the native \HII\ region and the 1~kpc rebinned spaxel results.

\section{Discussion}\label{sec:disucssion}
Given that gas-phase abundances of \HII\ regions are created by previous generations of stars, we can start to examine possible links between current and previous generation of star formation via the star formation history with the locally enriched ISM.

\subsection{Resolved Metal Enrichment and Connection to the Localized Physical Conditions of the ISM}\label{sec:discussion_1}
We observe a scatter in the relative abundance $\Delta$(O/H) within the \HII\ regions of $\lesssim$0.1~dex, significantly less than the typical metallicity radial gradient over the length of the entire galaxy disk. The strongest correlation in the relative abundance $\Delta$(O/H) occurs with the ionization parameter $U$ (Figure~\ref{fig:z_vs_q}). This positive correlation is still present but much weaker before the radial metallicity gradient is removed and the ionization parameter is compared directly with the 12+log(O/H) metallicity. While there is a slight negative correlation between 12+log(O/H) metallicity and pressure for every galaxy (Figure~\ref{fig:z_vs_pres}), this negative trend is removed once we remove the radial metallicity gradient and we recover a lack of any strong correlation between the relative abundance $\Delta$(O/H) with the ISM pressure. Given that the radial ionization parameter and the ISM pressure gradients are both remarkably flat (Figures~\ref{fig:qgrad} and \ref{fig:presgrad}), the observed (or lack of a) correlation simply cannot arise from correlated variables. Thus the localized enrichment at the scale of individual \HII\ regions correlates with the local ionization parameter of the ionizing stellar population but is not strongly dependent on the ISM pressure. 

The strong correlation we observe between metallicity and ionization parameter is in contrast with the expected anti-correlation based on prior studies, where low-metallicity galaxies and/or \HII\ regions typically are observed to have large ionization parameters \citep[e.g.,][]{dopita86, maier06}. However, recent resolved studies show the same positive trend between metallicity and ionization parameter that we find, in resolved regions of luminous infrared galaxies \citep{dopita14} and resolved observations of \HII\ regions in normal star-forming disk galaxies \citep{kreckel19}. The breakdown of the metallicity-ionization parameter relation showing different results for spatially resolved data compared to global observations is not unexpected as resolved information embedded in individual \HII\ regions can be washed out when averaged over entire galaxies. Systematics can also arise due to comparing different galaxies with differing global metallicities, where the offset in metallicity is established by the stellar mass of that system \citep[mass-metallicity relation;][]{tremonti04}.

\citet{dopita14} put forward a geometry-related effect that drives the observed positive correlation where active star-forming regions have a different distribution of molecular gas, favoring higher ionization parameters for regions of high star formation, and thus, correlates with localized regions with enhanced abundances. Much like the findings in \citet{kreckel19}, in general we also do not find that more enriched regions are preferentially distributed in the spiral arms of the galaxies (Figure~\ref{fig:hiiregions_metallicity_n2o2}). Because the \HII\ regions are located uniformly throughout their galactic disk, the local ISM chemical enrichment cannot solely be driven by the immediate \HII\ region environment due to the lack of observed systematic trends with galactic environment and features. We find that the local ISM chemical enrichment is best traced by ionization parameter, driven by the amplitude and shape of the ionizing sources, the ISM density, and \HII\ region geometry. 

The ISM pressure, on the other hand, is not as straight forward to interpret, primarily due to the complex density structures of \HII\ regions. In an expanding bubble model \citep[or radiation-driven implosion scenario;][]{bertoldi89}, the pressure of the surrounding ISM is related to both the temperature and density of the ISM where the \HII\ region bubble is powered by the central stellar ionizing source with the shock front expanding due to the thermal pressure that accompanies the ionization of the surrounding gas. The radiation from the central hot stars penetrates the ISM where the flow of the ionized gas streams ultimately into the lower density surrounding ISM, heating the cold, low-density gas. This heating amplifies over densities within the ISM and results in further star formation \citep{dale12a}. This triggering of star formation from radiation-driven models creates complex geometric substructures in the turbulent patches of dense gas \citep{tremblin13, schneider16} or champagne outflows driven and other complex geometries that are stellar wind driven \citep{krumholz09, park10, silich13}. In the ``collect and collapse'' model \citep{elmegreen95}, the expansion of an \HII\ region into a supersonic turbulent cloud causes coagulation of clumps that are gravitationally unstable. These unstable clumps are then able to collapse further and form new stars. This is further complicated by observed anti-correlations between the electron temperature and the gas density \citep[e.g.,][]{spinoglio15}. In a clumpy nebula with high-density gas clumps, the clumps are likely to be unresolved by typical IFS surveys and highlight the difficulty in constraining the relative impact of pressure on local ISM conditions and unravelling the impact of density and temperature on the pressure models. 

An important caveat to keep in mind is that star formation occurs in a clustered, hierarchical nature of up to several hundred parsecs that are also correlated in age \citep{elmegreen09a}. This suggests that large star-forming complexes in the disk of galaxies are capable of undergoing extended and continuous star formation for $\sim$10~Myr in nebulae and tens of Myr in correlated star-forming complexes \citep{grasha17b, rahner18}. Therefore, the observed locally enriched trends may simply arise from efficient mixing on the scale of individual \HII\ regions. Such mixing is possible with sustained and continuous star formation occurring within star-forming complexes where \HII\ regions reside on scales of less than a few hundred parsecs. Due to rather inefficient mixing on kiloparsec scales in galaxies \citep{deavillez02}, it is entirely reasonable for multiple generations of star formation to occur before galaxy mixing processes can fully mix the enhanced abundances throughout the ISM. This would lead to pockets of enhanced relative metal abundances $\Delta$(O/H) at resolved \HII\ region resolutions.

\subsection{Spatial Resolution and Un-resolved ISM Properties}\label{sec:discussion_2}
Measurements of the radial metallicity gradient (Figure~\ref{fig:metgrad_n2o2_1kpc}) and absolute values in the ionization parameter and pressure (Figures~\ref{fig:qgrad_1kpc} and \ref{fig:presgrad_1kpc}, respectively) can deviate significantly from the true gradient and values as the spatial resolution is decreased. We do point out, however, as demonstrated by \citet{poetrodjojo19}, contamination by the DIG may have a more significant effect on the flattening of metallicity gradients than spatial smoothing. 

The ionizing radiation produced by the star clusters is only partially absorbed by the \HII\ regions. Superbubble models are believed to create a complex density and ionization structure that can be porous to ionizing radiation, allowing some radiation to escape \citep{shields90}. The gas that receives this leaked ionizing radiation that escaped from nearby \HII\ regions is the DIG. The radiation in the DIG is believed primarily to be photoionization by the radiation from massive stars \citep{martin97, oey07}. However, the DIG but may also be powered by shock excitation \citep{ramirez-ballinas14} or dust-scattered radiation \citep{barnes15}. DIG produced by evolved post-AGB stars are characterized by very high temperatures \citep{zhang17}. As demonstrated by \citet{poetrodjojo19}, distinguishing and removing DIG is an important step to accurately measure radial metallicity gradients. However, this step is beyond the scope of this immediate paper. We do not calculate the relative contribution of spatial smoothing vs DIG contamination toward the flattening of metallicity gradients.

We are confident that DIG contamination or spatial smoothing do not impact our radial fits for metallicity, ionization parameter, or pressure (Table~\ref{tab:gradients_hii}), because we resolve individual \HII\ regions and use spaxels consistent with pure photoionization only. 
However, this is likely not the case when we artificially degrade the spatial resolution of our data cubes to simulate higher redshift, lower spatial resolution observations with 1~kpc spaxels (Figures~\ref{fig:metgrad_n2o2_1kpc}, \ref{fig:qgrad_1kpc}, and \ref{fig:presgrad_1kpc}). The choice of 1~kpc spaxel size was selected to aid in comparison of the typical spatial scale of other large IFU surveys such as MaNGA and SAMI. 

The biggest uncertainty posed by unresolved or global ISM pressure or density measurements is the difficulty to model the spectra from an average of multiple \HII\ regions \citep{kewley19_araa}, and constrain the impact of DIG contamination, if any, on the ISM properties. We do find a strong dependence of the relative offset of pressure and ionization parameters with spatial resolution; observations where we are unable to resolve sub-\HII\ region scales have higher ionization parameter and pressure values compared to the native sub-\HII\ region resolution datacubes. We lack available wavelength coverage to observe multiple line ratios that cover a range of critical densities to fully exploit the dependence of the measured pressure in individual or unresolved \HII\ regions on spatial resolution. While we demonstrate that limited spatial resolution impacts the recovered local ISM properties, future high-resolution observations with multiple species and ionization states will enable sampling the full set of pressures, and therefore modeling the integrated properties, across a galaxy.

\begin{deluxetable*}{lccccc}
\tabletypesize{\scriptsize}
\tablecaption{Spearman $\rho$ rank correlation coefficients \label{tab:4}}
\tablecolumns{6}
\tablewidth{0pt}
\tablehead{
\colhead{} & 
\multicolumn{2}{c}{Figure~\ref{fig:z_vs_q}} &
\colhead{} & 
\multicolumn{2}{c}{Figure~\ref{fig:z_vs_pres}}
\\		
 \cline{2-3} \cline{5-6}\noalign{\smallskip} 
\colhead{Galaxy} & 
\colhead{12+log(O/H) vs. $\log U$} & 
\colhead{$\Delta$(O/H) vs. $\log U$} &
\colhead{} & 
\colhead{12+log(O/H) vs. log(P/k)} &  
\colhead{$\Delta$(O/H) vs. log(P/k)}   
}
\startdata 
NGC 1566	& $\rho$ = 0.23 & $\rho$ = 0.48 && $\rho$ = --0.44	& $\rho$ = --0.21 \\
            & p = 0.0029    & p = 0.0000061 && p = 0.000025	    & p = 0.020 \\
NGC 2835	& $\rho$ = 0.17 & $\rho$ = 0.36 && $\rho$ = --0.20	& $\rho$ = 0.06 \\
            & p = 0.055     & p = 0.000028  && p = 0.12	        & p = 0.66 \\
NGC 3521	& $\rho$ = 0.18 & $\rho$ = 0.30 && $\rho$ = --0.17	& $\rho$ = --0.20 \\
            & p = 0.089     & p = 0.0036    && p = 0.17	        & p = 0.093 \\
NGC 5068	& $\rho$ = 0.16	& $\rho$ = 0.26 && $\rho$ = --0.09  & $\rho$ = 0.17 \\
            & p = 0.073     & p = 0.0028    && p = 0.46	        & p = 0.15 \\
NGC 5236	& $\rho$ = 0.37	& $\rho$ = 0.21 && $\rho$ = --0.05	& $\rho$ = --0.06 \\
            & p = 0.000042  & p = 0.0077	&& p = 0.46	        & p = 0.51 \\
NGC 7793	& $\rho$ = 0.14	& $\rho$ = 0.33 && $\rho$ = --0.12  & $\rho$ = --0.08 \\
            & p = 0.025     & p = 0.0000015	&& p = 0.55	        & p = 0.33
\enddata
\tablecomments{
Columns list the Spearman $\rho$ rank correlation coefficients for the galaxies in this study. Below the correlation coefficients the p-values are listed for a hypothesis test whose null hypothesis is that two sets of data are uncorrelated. For all galaxies there is high statistical significance for the correlation between $\Delta$(O/H) and ionization parameter $U$. 
}
\end{deluxetable*}

\section{Summary and Conclusions}\label{sec:conclusion}
We investigate the connection between local physical conditions and their linkage to the localized enrichment of the ISM across the entire star-forming disks of six spiral galaxies using the 3D datacubes constructed in the TYPHOON Program. The full wavelength coverage from 3500-7000~\AA\ provides useful measures of the local chemical abundance, ionization parameter, and ISM gas pressure. The quality of the TYPHOON data allow us to cleanly separate and define individual \HII\ regions within the pseudo-IFU data cubes and measure line fluxes of each individual \HII\ region. The \HII\ regions in each galaxy are identified from the H$\alpha$ map using the \HII\ region finder algorithm \texttt{HIIphot} \citep{thilker00}. For each \HII\ region, we derive the oxygen abundance using the \n2o2\ metallicity diagnostic \citep{kewley19_araa}, the ionization parameter using the O32 diagnostic \citep{kewley19_araa}, and the ISM pressure diagnostic using the [SII]6717/[SII]6730 line ratios \citep{kewley19_apj}. 

The main findings of our study are summarized below.

\begin{enumerate}\itemsep4pt

\item All galaxies show a negative relation between the gas-phase oxygen abundance and galactocentric distance. All metallicity radial profiles are well-approximated with a simple linear metallicity gradient with small (mean 0.03~dex) scatter (Figure~\ref{fig:metgrad_n202}). NGC~1566 is an exception, with significant scatter and a hint of flattening of the radial metallicity profile beyond 15~kpc from the galactic center.
	
\item We recover flat profiles with galactocentric distance for both the radial ionization parameter (Figure~\ref{fig:qgrad}) and the radial ISM pressure (Figure~\ref{fig:presgrad}). 
    
\item Due to lack of strong spiral arms in the galaxies, we find marginal or no evidence of azimuthal metallicity variations (Figure~\ref{fig:delOH}). Because metal-enriched \HII\ regions are distributed throughout the disk, rather than preferentially residing within the spiral arms, there is no noticeable systematic trend of localized enrichment with galaxy environment.
    
\item We find a positive and remarkably tight correlation between the ionization parameter $U$ and the local chemical enrichment as measured by the residual metallicity relative to the radial trend $\Delta$(O/H) (Figure~\ref{fig:z_vs_q}). This is not a result of correlated observables, due to the lack of an observed ionization parameter radial gradient. We observe no trend between the relative metallicity $\Delta$(O/H) and pressure (Figure~\ref{fig:z_vs_pres}). The strong correlation between the ionization parameter and the relative metal abundance indicates that the physical conditions in \HII\ regions correlate with the local changes in metallicity. 
    
\item To test how the ionization parameter and pressure gradients are impacted by loss in spatial resolution, we resample our datacubes to simulate unresolved \HII\ regions at 1~kpc spaxel size. We find that the ionization parameter (Figure~\ref{fig:qgrad_1kpc}) and ISM pressure (Figure~\ref{fig:presgrad_1kpc}) gradients derived for the 1~kpc spaxels remain relatively flat and unchanged. However, the absolute value for the derived ionization parameter increases with decreasing resolution. ISM pressure values for the binned data are marginally larger relative to that from the native resolution. However, the radial fits are consistent within the uncertainties. 

\end{enumerate}

\acknowledgements
We are grateful for the enlightening discussions and valuable comments on this work by an anonymous referee that improved the scientific outcome and quality of the paper.
This paper is based on spectrophotometric data cubes obtained with the du~Pont 2.5m Telescope at the Las~Campanas Observatory \citep{bowen73}, in Chile, as part of the TYPHOON Program, which has been obtaining optical data cubes for the largest angular-sized galaxies in the southern hemisphere. We thank past and present Directors of the Carnegie Observatories (Drs. Wendy Freedman and John Mulchaey, respectively) and the numerous time assignment committees for their generous and unfailing support of this long-term program. 
K.G. is supported by the Australian Research Council through the Discovery Early Career Researcher Award (DECRA) Fellowship DE220100766 funded by the Australian Government. 
K.G. is supported by the Australian Research Council Centre of Excellence for All Sky Astrophysics in 3 Dimensions (ASTRO~3D), through project number CE170100013. 
K.G. also acknowledges support from Lisa Kewley's ARC Laureate Fellowship (FL150100113). 
This research has made use of the NASA/IPAC Extragalactic Database (NED) which is operated by the Jet Propulsion Laboratory, California Institute of Technology, under contract with NASA. 
This research also made use of NASA's Astrophysics Data System Bibliographic Services.
This research made use of Astropy,\footnote{http://www.astropy.org} a community-developed core Python package for Astronomy \citep{astropy13, astropy18}. 
Parts of the results in this work make use of the colormaps in the CMasher package \citep{vandervelden2020}.
K.G. appreciates the useful discussions and help on this work by T. Nordlander.
The authors thank the invaluable labor of the maintenance and clerical staff at their respective institutions, whose contributions make scientific discoveries a reality. This research was conducted on Ngunnawal Indigenous land. 

\facility{2.5m du Pont Telescope (Wide Field reimaging CCD (WFCCD) imaging spectrograph)}

\software{Astropy \citep{astropy13, astropy18}, 
CMasher \citep{vandervelden2020}; 
iPython \citep{ipython}, 
Matplotlib \citep{matplotlib}, 
Numpy \citep{numpy11, numpy20}, 
HIIphot \citep{thilker00},
pPXF \citep{cappellari04, cappellari17},
LZIFU \citep{ho16}.}

\bibliography{ms_metgradient_nz_apj_accepted_arxiv.bbl}

\end{CJK*}
\end{document}